\documentclass[11pt]{iopart}
\usepackage{datetime}
\usepackage{iopams}
\usepackage{epsf}
\usepackage{epsfig}
\usepackage{color}
\usepackage{psfrag}
\usepackage{verbatim}
\usepackage{setstack}
\usepackage{amsopn}
\usepackage{appendix}
\usepackage{graphicx}
\usepackage{graphics}
\usepackage{subfig}
\usepackage[margin=3.3cm]{geometry}
\usepackage[font=footnotesize,labelfont=bf]{caption}

\newcommand{\be}{\begin{eqnarray}}	
\newcommand{\ee}{\end{eqnarray}}

\newcommand{\comments}[1]{}   

\begin{document}

\eqnobysec

\title{Percolation and coarsening in the bidimensional voter model}
\author{Alessandro Tartaglia,  Leticia F. Cugliandolo and \\
Marco Picco
}

\address{Sorbonne Universit\'es, Universit\'e Pierre et Marie Curie -- Paris VI, \\
Laboratoire de Physique Th\'eorique et Hautes Energies,\\
4 Place Jussieu, 
75252 Paris Cedex 05, France
}


\begin{abstract}
We study the  bidimensional voter model on a square lattice with numerical simulations.
We demonstrate that the evolution takes place in two distinct dynamic regimes; a first
approach towards critical  site percolation and a further approach towards 
full consensus. We calculate the time-dependence of the two growing lengths finding that they are both algebraic
though with different exponents (apart from possible logarithmic corrections). We analyse the 
morphology and statistics of clusters of voters with the same opinion. We compare these results 
to the ones for curvature driven two-dimensional coarsening.  
\end{abstract}

\tableofcontents
\newpage

\section{Introduction}
\label{sec:intro}

Purely dynamical stochastic models are used to describe problems 
beyond physics such as opinion formation~\cite{Castellano09} and population genetics~\cite{TilmanKareiva97},
and treat issues in ecology, linguistics, etc. In the former context, questions on  the spatial spreading
of opinions are posed in terms of coarsening or segregation, just as in physical materials.

The voter model~\cite{CliffordSudbury73,HolleyLiggett75,Liggett99}
is one such purely dynamical stochastic model, used to describe
the kinetics of catalytic reactions~\cite{Krapivsky92,Krapivsky92b,FrachebourgKrapivsky96} and as a prototype model of 
opinion dynamics~\cite{Vazquez03,FernandezGracia14}. 
In its simplest realisation a bi-valued opinion variable, $s_i=\pm 1$, is 
assigned to each site on a  lattice or graph with some procedure that determines the 
initial conditions. Typically, the initial state is taken to be unbiased, with equal number of one and the other 
state. The dynamic rule is straightforward: 
at each time step, a variable chosen at random  adopts the opinion of a randomly-chosen neighbour. 
These moves mimic the influence of the neighbourhood on the individual opinion. The probability of the chosen spin to 
flip in a time step is simply given by the fraction of neighbours with opposite orientation.
The model is parameter free and invariant under global inversion of the spins, that is to say,
${\mathbb Z}_2$ symmetric. As a site surrounded by others sharing the same
opinion cannot fluctuate, there is no bulk noise and the dynamics are uniquely 
driven by interfacial noise. In some papers the model is defined in terms of a site-occupation variable instead of a spin.

Mathematicians, more precisely probabilists, solve this model by using a mapping to 
random walk theory~\cite{CliffordSudbury73,HolleyLiggett75}.
Physicists, instead, treat it within the master equation formalism. Once written in this form,
one reckons that the transition probabilities do not satisfy detailed balance and, therefore, the 
model is essentially out of equilibrium. Even though there is no asymptotic thermal state, 
the dynamics can be solved  as Glauber did for the stochastic Ising chain
since the equations for the correlations of different order decouple~\cite{Krapivsky92,FrachebourgKrapivsky96,ScheucherSpohn88}.

The voter model's evolution shows spatial clustering of similar 
opinions. It approaches one of the  two absorbing states with complete consensus
via a coarsening process in $d\leq 2$. It may also approach consensus in finite size $d>2$  systems but
only because of a large random fluctuation  with some 
small probability (that vanishes in the infinite size limit). Otherwise, 
an infinite family of disordered steady states exist in $d> 2$~\cite{ScheucherSpohn88,CoxGriffeaths86}.
The coarsening process in $d=2$ is very different from the curvature driven one, 
as can be appreciated in Fig.~\ref{fig:snapshots} where a series of snapshots of the 
spin configuration at increasing times are shown,
proving that the long-term dynamics are not determined by symmetry properties alone. It is also different from critical relaxation,
specially because of the lack of bulk fluctuations, see also Fig.~\ref{fig:snapshots}.

The absence of bulk noise and surface tension entail important differences with respect to curvature 
driven phase ordering kinetics~\cite{Bray94,Puri09,Cu15}. In the voter model, regions of one opinion can only be penetrated by the other at the boundary.
Besides, a large bubble consisting of voters of the same opinion does not shrink as in curvature driven processes.  
It slowly disintegrates as its boundary roughens diffusively to reach a
typical width of the order of the initial radius~\cite{Dornic-etal01,DellAsta06}, while the radius of the droplet remains 
statistically constant (the radially-averaged magnetisation profiles have a stationary middle point). 

Coarsening processes  usually conform to the dynamic scaling 
hypothesis~\cite{Bray94,Puri09,Cu15}. This assumption states that if there is a single growing length in the process, say $\ell(t)$, the statistical properties 
of the system are self-similar with respect to it. Under this assumption the space-time correlation is 
independent of time when distance is rescaled by $\ell(t)$.
In the voter model the evolution of a random initial condition shows the growth of ordered spatial regions. However,
the exact asymptotic solution of an {\it infinite system} in $d=2$ exhibits 
logarithmic violations of the standard scaling forms~\cite{Bray94}. Although a characteristic length 
$\ell(t) \simeq \sqrt{t}$ can be identified, 
the density of interfaces decays as $1/\ln t$ and the 
scaling function for the space-time correlation function $C(r, t)$
involves an additional logarithmically decaying factor~\cite{ScheucherSpohn88} (somehow similarly to the critical dynamic scaling~\cite{Hoha77,Caga05}
though with a logarithm instead of an algebraic correction).

The goal of  this work is to  characterise  the coarsening process in the bidimensional voter model with large but 
{\it finite linear size} by studying,
in detail, the geometric and statistical properties of the dynamic pattern of domains. Following the analysis in~\cite{ArBrCuSi07,SiArBrCu07,Blanchard,BlCoCuPi14} 
we will demonstrate that the system evolves in two time regimes: a pre-asymptotic approach to critical percolation and an
ultimate approach to full consensus. With the aim of identifying and distinguishing the growing length
in each of  these regimes, we compute special time-dependent observables such as the number density 
of domains with a given area, or the number density of interfaces with a given length. As the characteristic length 
associated to the approach to percolation, that we call $\ell_P(t)$, grows quite slowly in time 
we are able to analyse this dynamic regime in detail (contrary to the what happens in the Ising model, where $\ell_P(t)$ 
is such a fast growing function of time that in practice critical percolation is reached too quickly to allow for a 
careful study of this dynamic regime).

The paper is organised as follows. 
In Sec.~\ref{sec:analytic} we introduce the model and we summarise the time-dependence of several observables 
that were derived analytically in the past for infinite size systems. 
In Sec.~\ref{sec:numerical}  we present our numerical results. 
We discuss the violation of the scaling hypothesis as observed in the time-dependence of the density of interfaces, 
persistence probability, two-time correlation function and space-time correlation function of an infinite size system. We
then show our novel results  on geometric properties of the largest cluster, number density of domain areas and interface lengths in finite 
size systems. We relate their properties to the 
fractal properties of these objects. More importantly, this analysis allows us to demonstrate the existence 
of the two dynamic regimes evoked in the previous paragraph: a first approach to critical site percolation and the further 
evolution towards complete consensus in a longer time-scale. We end the paper with a concluding section.

\section{Analytical results}
\label{sec:analytic}

The definition of the voter model is extremely simple. Each node $i$ of a graph is endowed with a binary variable 
$s_i=\pm 1$. At each time step an agent $i$ is selected at random along with one of its neighbours $j$
and the selected agent takes the opinion of the neighbour, i.e $s_i=s_j$. In the case of a voter model on a
$d$-dimensional hypercubic lattice, the spin-flip rate for the site $\boldsymbol{x}$ is given by~\footnote{Frachebourg and 
Krapivsky~\cite{FrachebourgKrapivsky96} define the spin-flip rate
with a factor $1/\tau$ in front of the parenthesis, they take  $ \tau = 4/d $, and they therefore have an overall factor $d/4$. 
This coincides with our definition of $W_{\boldsymbol{x}}$ since  we have a factor $1/(2\tau)$, we choose $\tau d=2$ and we  
also have a factor $d/4$ overall. There is, however,  a difference with the choice made by Ben-Naim \textit{et al.}
who used a  $ \tau $ that is half ours in their calculations~\cite{BenNaim-etal96}. 
}
\begin{equation}
W_{\boldsymbol{x}}(s) = \frac{1}{2\tau} \left(1-\frac{1}{2d} \ s_{\boldsymbol{x}}(t)
\sum_{ \boldsymbol{y}\in \mathcal{N}(\boldsymbol{x}) } s_{\boldsymbol{y}}(t) \right)
\;  ,
\label{spin_flip_rate}
\end{equation}
where $s={\left( s_{\boldsymbol{x}} \right)}_{ \boldsymbol{x} \in \mathbb{Z}^d}$ denotes the state of the system
at time $ t $,
$s_{\boldsymbol{x}} $ is the value of the spin on site $\boldsymbol{x}$, 
$\mathcal{N}(\boldsymbol{x})$ is the set of
its neighbouring sites, and $\tau$ defines the timescale of the process. This particular form
of spin-flip rate, which is just a constant times the fraction of disagreeing neighbouring sites, defines the 
 so-called linear voter model. It is possible to define other voter-like models in which the spin-flip rate is not simply
a linear function of the local effective field 
$h_{\boldsymbol{x}}=\sum_{ \boldsymbol{y}\in \mathcal{N}(\boldsymbol{x}) } s_{\boldsymbol{y}} $,  but still satisfy
the $ \mathrm{Z}_2 $ symmetry and have similar properties~\cite{DrouffeGodreche99, Castellano09}. We will 
focus on the model with spin-flip rate (\ref{spin_flip_rate}) here. Note also that we are taking $\boldsymbol{x} \in \mathbb{Z}^d$.
We will assume the lattice to be infinite in all the calculations appearing further in this section, 
though we will be especially concerned with finite size effects in Sec.~\ref{sec:numerical}.

\par
Equation~(\ref{spin_flip_rate}) implies that this spin model has no bulk noise, {\it i.e.}, if a site `agrees' with 
all its nearest-neighbours, its spin-flip rate vanishes. In this 
sense, the dynamics are similar to the zero-temperature Glauber ones. The consequence is that the `consensus'  states, {\it {\it i.e.}},
the states  in which all sites have the same opinion, are `absorbing' states. Indeed, if the system reaches one of the two
consensus states, it will never leave it. 

\par
However, this does not mean that the asymptotic steady state  must be
one of full consensus. In fact, it turns out that the coarsening process is not always effective in bringing the system towards
a single-domain state, and whether it does or not depends on the dimensionality of the lattice. For $ d\le2 $ the system coarsens 
until ultimately reaching a single domain state, while for $ d>2$ there is an infinite family of non-completely-ordered steady 
states~\cite{ScheucherSpohn88, CoxGriffeaths86}. The discrepancy in the asymptotic regime reached  above and 
below $d=2$ will be further discussed in this section.

\par
The probability distribution satisfies the master equation
\begin{equation}
 \frac{ \mathrm{d}}{\mathrm{d}t} P(s,t) = 
 \sum_{\boldsymbol{x}} \Big( W_{\boldsymbol{x}}(s^{\boldsymbol{x}}) 
  P(s^{\boldsymbol{x}},t) - W_{\boldsymbol{x}}(s) P(s,t) \Big)
  \; , 
\label{master_eq}
\end{equation}
where $ s^{\boldsymbol{x}} $ is the configuration that differs from  $ s $ only
in that the spin on the site $ \boldsymbol{x} $ is reversed. One can then derive a set of differential 
equations for the $n$-spin time-dependent correlation functions $ {\langle s_{\boldsymbol{x_1}}...s_{\boldsymbol{x_n}} \rangle}
=\sum_{s} s_{\boldsymbol{x_1}}...s_{\boldsymbol{x_n}} P(s,t) $ and find that, since the update rule $W_{\boldsymbol{x}}$ is simply
linear in the local spin, the equations for the correlation functions of different order decouple. 

The single-body
correlation function or average magnetisation satisfies~\cite{FrachebourgKrapivsky96, BenNaim-etal96}
\begin{equation}
  \frac{ \mathrm{d}}{\mathrm{d}t} \langle s_{\boldsymbol{x}} \rangle = -2 \langle s_{\boldsymbol{x}} W_{\boldsymbol{x}}(s)
  \rangle = \frac{1}{2 \tau d} \Delta_{\boldsymbol{x}} \langle s_{\boldsymbol{x}} \rangle
  \; , 
\label{magnetization}
\end{equation}
where $ \Delta_{\boldsymbol{x}}  $ denotes the discrete Laplace operator, 
\begin{equation}
  \Delta_{\boldsymbol{x}} \langle s_{\boldsymbol{x}} \rangle \equiv -2 d \langle s_{\boldsymbol{x}} \rangle +
  \sum_{i=1}^d \Big( \langle s_{ \boldsymbol{x}+\boldsymbol{\mathrm{e}}_i } \rangle +
  \langle s_{ \boldsymbol{x}-\boldsymbol{\mathrm{e}}_i } \rangle \Big)
  \; , 
\label{Laplace_op}
\end{equation}
and $ \{ \boldsymbol{\mathrm{e}}_i \}_{i=1,...,d} $ are the set of unit vectors defining the lattice. In the infinite system size 
limit or for periodic boundary conditions all sites satisfy this same equation.
For finite size systems with open boundary conditions the  sites at the edges should be considered separately.

Similarly, for the two-body correlation function one has ~\cite{FrachebourgKrapivsky96, BenNaim-etal96}
\begin{equation}
  \frac{ \mathrm{d}}{\mathrm{d}t} \langle s_{\boldsymbol{x}} s_{\boldsymbol{y}} \rangle = 
  -2 \Big \langle s_{\boldsymbol{x}} s_{\boldsymbol{y}} \Big( W_{\boldsymbol{x}}(s) +
  W_{\boldsymbol{y}}(s) \Big) \Big \rangle = 
\frac{1}{2 \tau d} \Big( \Delta_{\boldsymbol{x}} + \Delta_{\boldsymbol{y}} \Big) \langle s_{\boldsymbol{x}} s_{\boldsymbol{y}} \rangle 
\; . 
\label{2body_correlf}
\end{equation}

Interestingly enough, Eq.~(\ref{master_eq}) is mathematically equivalent to the master
equation for a continuous-time symmetric random walk on $ \mathbb{Z}^d $ with jumping rate $\tau^{-1}$.
As a result, the mean magnetisation per site, defined as $ m(t) = \lim_{L \rightarrow \infty}
L^{-d} \sum_{\boldsymbol{x} \in \{1,...,L\}^d} \langle s_{\boldsymbol{x}}(t) \rangle $ 
plays the role of the total probability for the walker and is thus a conserved quantity. The same result can be obtained
by summing both sides of Eq.~(\ref{magnetization}) over all lattice sites. 
Notice that while the magnetisation of a specific system does change in a single update event, 
the average over all sites and over all trajectories of the dynamics is conserved. 

\par
Consider a finite system with an initial fraction $ \rho $ of voters in the $ +1 $ state and  $1-\rho$
in the $ -1 $ state, so that the initial magnetisation density is  $ m_0=2\rho -1 $. Suppose that the system reaches
consensus in which the state of magnetisation $ m=+1 $ occurs with probability $ p_{+1}(\rho) $ and the state with $ m=-1 $
with probability $ 1 - p_{+1}(\rho) $. Then since $ m_0 = m_{\infty} $ one has $ 2\rho -1 = 2 p_{+1}(\rho) -1$, hence
$ p_{+1}(\rho) = \rho $.

\par
Concerning again Eq.~(\ref{magnetization}), by using the discrete Fourier transform of 
$ \langle s_{\boldsymbol{x}} \rangle$, one can prove that its general solution on an infinite 
size lattice has the form~\cite{FrachebourgKrapivsky96,BenNaim-etal96}
\begin{equation}
  \langle s_{\boldsymbol{x}}(t) \rangle = \mathrm{e}^{-\frac{t}{\tau}} \sum_{ \boldsymbol{y} \in \mathbb{Z}^d}  
  \sigma_{\boldsymbol{y}} \, \mathcal{J}_{\boldsymbol{x}-\boldsymbol{y}} \left( \frac{t}{\tau d} \right)
\label{mag_sol} 
\end{equation}
where $ \sigma_{\boldsymbol{y}} = \langle s_{\boldsymbol{y}}(t=0) \rangle$ and
$ \mathcal{J}_{\boldsymbol{x}} $ is a shorthand notation for the multi-index modified Bessel functions,
$ \mathcal{J}_{\boldsymbol{x}} (u) = \prod_{i=1}^d \mathcal{I}_{x_i} (u)$, with $ \mathcal{I}_{ \alpha } $
the usual modified Bessel function of order $ \alpha $.

If the initial configuration  is such that a single $ +1 $ voter sits at the origin and is surrounded by a ``sea'' of undecided
voters ({\it i.e.} $ s_{\boldsymbol{x}} (0) = \pm 1\ $ with probability $1/2$, for all $\boldsymbol{x} \neq \boldsymbol{0} $) then, since
$ \sigma_{\boldsymbol{x}} = \delta_{\boldsymbol{x},\boldsymbol{0}}$, the solution to Eq.~(\ref{magnetization}) reduces to 
$ \langle s_{\boldsymbol{x}}(t) \rangle = \mathrm{e}^{-t / \tau} \mathcal{J}_{\boldsymbol{x}} ( t/(\tau d) ) $.
By using now the asymptotic relation $ \mathcal{I}_{\alpha} (z) \sim \mathrm{e}^z / \sqrt{2 \pi z} $, $z\gg 1$, valid for any real $ \alpha $,  
one  finds the asymptotic behaviour of the average site magnetisation,
$ \langle s_{\boldsymbol{x}}(t) \rangle \sim [ 2 \pi t/ (\tau d) ]^{-d/2} $. Thus, a single voter relaxes to the average 
undecided opinion of the rest of the population.

\par
The last result is exact, but does not provide meaningful information on how the steady state of the system is reached.
In this sense, a more interesting quantity  is the two-body correlation function determined by Eq.~(\ref{2body_correlf}). In 
order to solve this equation~\cite{FrachebourgKrapivsky96} one makes the assumption
that at each time $ t $ the state of the system is translationally invariant, so that
$ \langle s_{\boldsymbol{x}} s_{\boldsymbol{y}} \rangle $ depends on the lattice vectors $ \boldsymbol{x} $ and $ \boldsymbol{y} $ 
only through their difference $ \boldsymbol{n} = \boldsymbol{x} - \boldsymbol{y} $. 
Then,  by denoting $ \mathrm{G}_{\boldsymbol{n}} (t)  = \langle s_{\boldsymbol{x}} (t) s_{\boldsymbol{x}+\boldsymbol{n}} (t) \rangle $, 
Eq.~(\ref{2body_correlf}) simplifies to 
\begin{equation}
 \frac{ \mathrm{d}}{\mathrm{d}t} \mathrm{G}_{\boldsymbol{n}} (t) = 
\frac{1}{ \tau d} \, \Delta_{\boldsymbol{n}} \mathrm{G}_{\boldsymbol{n}} (t)
\label{correl_eq}
\end{equation}
which should be solved subject to the boundary condition $ \mathrm{G}_{\boldsymbol{0}} (t) = \langle s^2_{\boldsymbol{x}} (t) \rangle = 1 $, for any $ t $.
In addition, it is natural to choose the initial condition $ \mathrm{G}_{\boldsymbol{n}} (0) = \delta_{\boldsymbol{n},\boldsymbol{0}} $, that corresponds 
to a completely uncorrelated initial state. Equation (\ref{correl_eq}) is basically identical to Eq.~(\ref{magnetization}) apart from numerical factors, and
one would be tempted to consider a solution of the form $ \widetilde{\mathrm{G}}_{\boldsymbol{n}} (t) =  
 \mathrm{e}^{-\frac{2 t}{\tau}} \mathcal{J}_{\boldsymbol{n}} \left( 2 t/ (\tau d) \right) $. However, $ \widetilde{\mathrm{G}}_{\boldsymbol{0}} (t)$
does not satisfy the boundary condition. In order to maintain $ \mathrm{G}_{\boldsymbol{0}} (t) = 1 $ throughout the evolution,
one can reformulate the problem posed by  Eq.~(\ref{correl_eq}) as the equivalent lattice
diffusion problem with a constant localised source at the origin, and look for a solution of the form
\begin{equation}
 \mathrm{G}_{\boldsymbol{n}} (t) =  
 \mathrm{e}^{-\frac{2 t}{\tau}} \ \mathcal{J}_{\boldsymbol{n}} \left( \frac{2 t}{\tau d} \right) +
 \int^{t}_0 {   \mathrm{d} t^{\prime} \,  S_d(t - t^{\prime}) \, \mathrm{e}^{-\frac{2 t^{\prime}}{\tau}} \
 \mathcal{J}_{\boldsymbol{n}} \left( \frac{2 t^{\prime}}{\tau d} \right)    }
\label{correl_sol}
\end{equation}%
with $S_d(t) $  the ``strength'' of the source. From a physical point of view, this solution corresponds 
to placing a source $ \mathrm{G}_{\boldsymbol{0}}(t=0) $ at the initial time at the origin and supplement it by an additional input
$ S_d(t) dt $ that is added during the time interval $ \left( t, t+dt \right) $ to keep the overall value unchanged. Equation~(\ref{correl_sol}) evaluated at the
origin ($ \boldsymbol{n} = \boldsymbol{0} $) becomes 
\begin{equation}
 1  = \left[ \mathrm{e}^{-\frac{2 t}{\tau d }} \ \mathcal{I}_0 \left( \frac{2 t}{\tau d} \right) \right]^d +
 \int^{t}_0   \mathrm{d} t^{\prime} \,   S_d(t - t^{\prime}) 
 \left[ \mathrm{e}^{-\frac{2 t^{\prime}}{\tau d }} \ \mathcal{I}_0 \left( \frac{2 t^{\prime}}{\tau d} \right) \right]^d
 \; . 
\label{boundary_cond}
\end{equation}%
By using now the Laplace transform of the strength, $ \hat{S}_d(\lambda) = \int^{+\infty}_0 \mathrm{d}t \,  S_d(t) \mathrm{e}^{-\lambda t}$,
and the Laplace transform $ \hat{T}_d(\lambda) $ of the function $ T_d(t) = \left[ 
\mathcal{I}_0(t) \mathrm{e}^{-t} \right]^d $, one arrives at 
\begin{equation}
  \hat{S}_d(\lambda) = -1 + \frac{2}{ \tau d } \Big[ \lambda \cdot  \hat{T}_d \left( \frac{\tau d}{2} \lambda \right) \Big]^{-1}
  \; . 
\label{Laplace_transf}
\end{equation}%
Using now the integral representation of the modified Bessel function $ \mathcal{I}_0 $, namely
$ \mathcal{I}_0(x) = \frac{1}{2 \pi} \int^{2 \pi}_0 \mathrm{d}q \, \mathrm{e}^{x \cos(q)} $, it is possible
to express $ \hat{T}_d $ in terms of the Watson integrals, 
\begin{equation}
  \hat{T}_d(\lambda) 
  = 
  \frac{1}{(2 \pi)^d} 
  \int^{2 \pi}_0 \cdots \int^{2 \pi}_0 \mathrm{d} q_1 \cdots \mathrm{d}q_d 
  \: \frac{1}{ \lambda + d - \sum^d_{i=1} \cos q_i  } \; , 
\label{Watson_integral}
\end{equation}%
and find an expression for $ \hat{S}_d(\lambda) $. For example, in the case $ d= 1 $, $ \hat{S}_1(\lambda) = 
\sqrt{ (\lambda+2)/\lambda } $. More complicated expressions arise when $ d $ is larger and ultimately there is no closed-form
for them. Nevertheless, we are just interested in the asymptotic behaviour of the source strength $ S_d $, which in turn is given by 
the low-$\lambda$ limit of its Laplace transform~\cite{FrachebourgKrapivsky96}, 
\begin{equation}
      \hat{S}_d(\lambda) \sim \left\{
                \begin{array}{l l l}
                  \left( \frac{\tau }{2} \lambda \right)^{-\frac{1}{2}} & \quad \mbox{if} \;\; d=1 & \\
                  \left( \tau \lambda \right)^{-1} \ln^{-1} \left[ 1/(\tau \lambda) \right] & \quad \mbox{if} \;\; d=2 & \quad \mbox{as} \;\; \lambda \rightarrow 0 \\
                  \left( \frac{\tau d}{2} \lambda \right)^{-1} & \quad \mbox{if} \;\; d>2 &
                \end{array}
              \right.
\label{source_transf}
\end{equation} 
and thus
\begin{equation}
      S_d(t) \sim \left\{
                \begin{array}{l l l}
                  \left( \frac{2t}{\tau }  \right)^{-\frac{1}{2}} & \qquad\qquad \mbox{if} \;\; d=1 & \\
                  \ln^{-1}\left( \frac{t}{\tau	}  \right)  & \qquad\qquad \mbox{if} \;\; d=2 & \quad \mbox{as} \;\; t \rightarrow + \infty  \\
                  \mbox{const.} & \qquad\qquad \mbox{if} \;\; d>2 &
                \end{array}
              \right.
\label{source_asympt}
\end{equation}

In $d=2$, the long-time behaviour of the source strength in the integral is
$ S_2(t-t') \simeq 1/ \ln[(t-t')/\tau] \simeq 1/\ln(t/\tau)$. Using the asymptotic relations for $ \mathcal{I}_{\alpha}$, and calling $ \boldsymbol{n} = (n_1, n_2) $,
Eq.~(\ref{correl_sol}) implies 
\begin{equation}
 \mathrm{G}_{\boldsymbol{n}} (t) \simeq 
 \frac{1}{2 \pi t}  +
  \frac{c}{ \ln (t/\tau) } \ \int^{t}_0 {   \mathrm{d} t^{\prime}  \, 
  \mathrm{e}^{- \frac{2 t^{\prime}}{\tau} } \ \mathcal{I}_{n_1} \left( \frac{t^{\prime}}{\tau} \right) \mathcal{I}_{n_2} \left( \frac{t^{\prime}}{\tau} \right)   }
\label{correl_asympt1}
\end{equation}%
%
as $ t \rightarrow + \infty$ dropping corrections $\mathcal{O} ( t^{-2} )$, with $ c$ a numerical factor to be determined. 
Using the integral representation of the modified Bessel function $ \mathcal{I}_n$  for integer values of $ n$, that is
$ \mathcal{I}_{n} ( t ) = (2 \pi)^{-1} \int^{\pi}_{- \pi} \mathrm{d} k \; 
\mathrm{exp} \left[ t \cos k -i \, n \ k \right] $, Eq.~(\ref{correl_asympt1}) reduces to
\begin{equation}
 \mathrm{G}_{\boldsymbol{n}} (t) \simeq  \;
  \frac{c}{ \ln(t/\tau) } \frac{1}{(2 \pi)^2} \int^{\pi}_{-\pi} \!\! \mathrm{d} k_1 \; \int^{\pi}_{-\pi} \!\! \mathrm{d} k_2 \; 
  \mathrm{e}^{- i \boldsymbol{n} \cdot \boldsymbol{k} } \, \hat{f} \left( \boldsymbol{k}, t \right)\; 
  + \mathcal{O} \left( \frac{1}{t} \right)
\label{correl_asympt2}
\end{equation}%
where $ \boldsymbol{k} = (k_1, k_2 )$ and the function $ \hat{f} \left( \boldsymbol{k}, t \right) $ is given by
\begin{equation}
 \hat{f} \left( \boldsymbol{k}, t \right) = \tau\
 \frac{ 1 - \mathrm{e}^{- \frac{t}{\tau} ( 2 - \cos k_1 - \cos k_2 )} }{ 2 - \cos k_1 - \cos k_2 } 
\; .
\label{structure_fact_expr}
\end{equation}
%
Apart from a time-dependent prefactor, one can recognise in $\hat{f}$ the dynamical structure factor of the system, which is
defined as the lattice Fourier Transform of the space-time dependent correlation function,
\begin{equation}
 S(\boldsymbol{k},t) =  \frac{1}{\ln{(t/\tau)}} \, \hat{f} \left( \boldsymbol{k}, t \right) \; \propto \; \;
  \sum_{\boldsymbol{n} \in \mathbb{Z}^2} { \mathrm{G}_{\boldsymbol{n}} (t) \; \mathrm{e}^{i \boldsymbol{n} \cdot \boldsymbol{k}  } }.
\label{structure_fact_def}
\end{equation}%
In the limit $  |\boldsymbol{k} | \rightarrow 0$, $\hat{f}$ can be approximated as
$ \hat{f} \left( \boldsymbol{k}, t \right) \simeq 2 \ \tau  \ k^{-2} ( 1- \mathrm{e}^{-\frac{t}{2 \tau} \, k^2} )$, 
where $ k = | \boldsymbol{k} |$,
{\it {\it i.e.}} it becomes isotropic in $k$-space. Then the large-distance behaviour of the correlation function
is characterized by the scaling form
\begin{equation}
  \mathrm{G}_{\boldsymbol{n}} (t) \; \sim \; 
     \frac{1}{\ln{(t/\tau) }} \ f \left( \frac{ | \boldsymbol{n} | }{ \sqrt{t/ 2 \tau} } \right)
\label{correlf_asympt3}
\end{equation}%
where the scaling function $f $ is just given by the inverse Fourier Transform of $ \hat{f}$. Equation~(\ref{correlf_asympt3}) clearly displays
the emergence of a dynamical characteristic length $\ell (t) $ which scales as $ \sqrt{t}$, and the logarithmic violation of dynamic scaling.

\par
An interesting quantity that can be extracted from the two-body correlation function is the density of reactive interfaces $ \rho$, defined
as the average value of the fraction of unsatisfied bonds or, equivalently, the fraction of neighbouring voters with 
disagreeing opinions. This quantity is linked to the correlation function through the relation
\begin{equation}
 \rho(t) = \frac{1}{2} \left( 1 - \frac{1}{2 d} 
\sum^d_{i=1}  \left[ \mathrm{G}_{ \boldsymbol{\mathrm{e}}_i }(t) +  \mathrm{G}_{ \boldsymbol{\mathrm{-e}}_i }(t) \right] \right) =
  \frac{1}{2} \left( 1 - \mathrm{G}_{ \boldsymbol{\mathrm{e}}}(t) \right)
\label{rho_definition}
\end{equation}%
where $ \boldsymbol{\mathrm{e}}_i $ are the lattice unit vectors. Note that the sum over the nearest-neighbours can be lifted since the dynamics is isotropic along
the $ d $ principal directions of the lattice. From Eq.~(\ref{correl_sol}) evaluated at $ \boldsymbol{n} = (1,0, ... ,0) $ and
the fact that $ \mathrm{G}_{\boldsymbol{0}} \equiv 1  $,
one obtains
\begin{eqnarray}
 && \rho(t) = \frac{1}{2} \, \mathrm{e}^{- \frac{2 t}{\tau}} \, \mathcal{I}^{d-1}_0\left(\frac{2t}{\tau d}\right) \left[
  \mathcal{I}_0\left(\frac{2t}{\tau d}\right) - \mathcal{I}_1\left(\frac{2t}{\tau d}\right) \right] \; 
  \nonumber\\
  && 
  + \; \frac{1}{2}
  \int^{t}_0 \mathrm{d}u \, S_d(t - u) \, \mathrm{e}^{- \frac{2u}{\tau}} \, \mathcal{I}^{d-1}_0\left(\frac{2u}{\tau d}\right) \left[
  \mathcal{I}_0\left(\frac{2u}{\tau d}\right) - \mathcal{I}_1\left(\frac{2u}{\tau d}\right) \right] .
\label{rho_eq}
\end{eqnarray}%
Combining the latter equation with Eqs.~(\ref{source_asympt}) and the asymptotic relations 
$ \mathcal{I}_0(z) \simeq \mathcal{I}_1 (z) \simeq  e^z\left[
 1/ \sqrt{2 \pi z} + \mathcal{O} (z^{-3/2}) \right] $ and $ \mathcal{I}_0(z) - \mathcal{I}_1 (z) \simeq e^z \left[ 1 / \sqrt{8 \pi z^3} +
 \mathcal{O} (z^{-5/2}) \right]$, the asymptotic behaviour of the density of reactive interfaces is found to be
\begin{equation}
      \rho(t) \sim \left\{
                \begin{array}{l l l}
                  t^{-\frac{1}{2}} & \qquad\qquad \mbox{if} \;\; d=1 & \\
                  \ln^{-1} (t/\tau )  & \qquad\qquad \mbox{if} \;\; d=2 & \quad \mbox{as} \;\; t \rightarrow + \infty \\
                  a - b t^{-d/2} & \qquad\qquad \mbox{if} \;\; d>2 &
                \end{array}
              \right.
\label{rho_asympt}
\end{equation}

These results allow us to establish some conclusions on the coarsening process  in the voter model: 
in $ d \le 2 $ the probability that two voters at a given separation had opposite opinion
 vanishes asymptotically, no matter how much distant they are, and coarsening eventually leads to a single-domain final state. 
In $ d > 2$,  an infinite system reaches a dynamic frustrated state, where opposite-opinion voters coexist and 
continually evolve in such a way
that the average concentration of each type of voters remains fixed. Dimension $ d=2 $ is particular since it lies at the border between the two cases.
There is a coarsening process which brings the system towards the single-domain state, but it is very slow, since  the density of active interfaces vanishes
only as $ 1/ \ln (t/\tau) $. 

\par
As a last effort, we derive the two-time correlation function, defined as $ A_{\boldsymbol{x}}(t,t_0) = 
\langle s_{\boldsymbol{x}}(t) s_{\boldsymbol{x}}(t_0) \rangle $,
which is an interesting quantity to look at since it provides information on the typical timescale for the process to reach a steady state.
For fixed $ t_0 $ and $ \boldsymbol{x_0} \in \mathbb{Z}^d$, let us introduce the function $ F_{\boldsymbol{x}} ( t ; \boldsymbol{x_0},t_0 ) = \langle s_{\boldsymbol{x}}(t+t_0) s_{\boldsymbol{x_0}}(t_0) \rangle $ 
for any $ \boldsymbol{x} \in \mathbb{Z}^d $ and $ t \ge 0 $. Dropping for a moment the dependence of $ F $ on $ t_0 $ and $ \boldsymbol{x_0} $, it is easy to see that it satisfies the same
equation as the single-body correlation function, {\it i.e.} $  \frac{ \mathrm{d}}{\mathrm{d}t} F_{\boldsymbol{x}} (t) = \frac{1}{2 \tau d} \Delta_{\boldsymbol{x}}  F_{\boldsymbol{x}}(t) $,
apart from a factor $1/2$.
Thus $ F_{\boldsymbol{x}} (t) = \mathrm{e}^{-t/ \tau} \sum_{\boldsymbol{y}} f_{\boldsymbol{y}} \mathcal{J}_{\boldsymbol{x} -\boldsymbol{y}} (t/ \tau d) $
where $ f_{\boldsymbol{y}}(\boldsymbol{x_0},t_0)= \langle s_{\boldsymbol{y}} (t_0) s_{\boldsymbol{x_0} } (t_0)\rangle $. 
Then assuming that at each time the state of the system is spatially translational invariant and using 
$ A_{\boldsymbol{x}}(t,t_0) = F_{\boldsymbol{x}} ( t -t_0 ; \boldsymbol{x},t_0 ) $, one gets
\begin{eqnarray}
  A_{\boldsymbol{x}}(t,t_0) 
  &=&
   \; \mathrm{e}^{-(t-t_0)/ \tau} \sum_{\boldsymbol{n} \in \mathbb{Z}^d } 
   \mathrm{G}_{\boldsymbol{n}} (t_0) \
    \mathcal{J}_{\boldsymbol{n}} \left( \frac{t-t_0}{\tau d} \right) \; , 
\label{ac_gen_sol}
\end{eqnarray}%
with the dependence on $ \boldsymbol{x} $ disappearing consistently with the hypothesis of translational invariance. 
As a simple check we verify that setting $t=t_0$ in (\ref{ac_gen_sol}) we find $ \mathrm{G}_{\boldsymbol{0}} (t_0)$. 
Indeed, using  ${\mathcal J}_{\boldsymbol{n}}(0) = \prod_{i=1}^d {\mathcal I}_{x_i}(0) = 0$ for all 
${\boldsymbol{n}} \neq {\boldsymbol{0}}$ and ${\mathcal J}_{\boldsymbol{0}}(0) = \prod_{i=1}^d {\mathcal I}_{x_i=0}(0) =1$ this fact is verified.

In the particular case $ t_0 = 0 $, if the initial configuration
is completely uncorrelated, {\it i.e.} $ \mathrm{G}_{\boldsymbol{n}} (0) = \delta_{\boldsymbol{n}, {\boldsymbol{0}}} $, the solution reduces to 
\begin{equation}
 A_0(t) = A (t,t_0=0) =  \mathrm{e}^{-t / \tau} \left[
  \mathcal{I}_0 \left( \frac{t}{  \tau d} \right)
 \right]^d
\label{ac0}
\end{equation}
with asymptotic behaviour $ A_0(t) \sim \left[ 2 \pi t/ (\tau d)  \right]^{-d/2}$. 

In the limit $t\gg t_0$ one can use the asymptotic expansion of ${\mathcal J}_{\boldsymbol{n}}(u) =
\prod_{i=1}^d {\mathcal I}_{x_i}(u) \simeq [{\rm e}^{u}/\sqrt{2\pi u}]^d$ with $u = (t-t_0)/(\tau d)$ and, therefore,
\begin{eqnarray}
  \lim_{t\gg t_0} A_{\boldsymbol{x}}(t,t_0) &=& 
     \; [2\pi (t-t_0)/(\tau d)]^{-d/2}
     \sum_{\boldsymbol{n} \in \mathbb{Z}^d } 
   \mathrm{G}_{\boldsymbol{n}} (t_0) \ .
\label{asympt_two_time_corr}
\end{eqnarray}
The $t_0$-dependent last factor can be estimated as follows
\begin{eqnarray}
K(t_0) \equiv \sum_{\boldsymbol{n} \in \mathbb{Z}^d } 
   \mathrm{G}_{\boldsymbol{n}} (t_0) \mapsto 
   \int d^dx \ C( \boldsymbol{x}, t_0) = 2\pi \! \int dr \, r \, C(r,t_0) 
\end{eqnarray}
with $C(\boldsymbol{x},t)$ the space-time correlation function in the continuum space limit. Setting $d=2$ and 
using the scaling function for $C(r,t_0)$ expressed in Eq.~(\ref{correlf_asympt3}) 
\begin{eqnarray}
K(t_0) &=& 2\pi \! \int dr \, \frac{r}{\ln (t_0/\tau) } \, f\left( \frac{r}{\sqrt{t_0/\tau}} \right)
\propto  \frac{2\pi}{\ln (t_0/\tau) }  \, \frac{t_0}{\tau } \; . 
\end{eqnarray}
Going back to Eq.~(\ref{asympt_two_time_corr}) this implies
\begin{eqnarray}
\lim_{t\gg t_0} A_{\boldsymbol{x}}(t,t_0) 
\propto \frac{1}{\ln ( t_0/\tau) }  \ \ ( t/t_0-1 )^{-1}
\label{eq:asympt_final_two_time_corr}
\; .
\end{eqnarray} 
Further details on how to obtain the analytical results sketched in this section can be found
in~\cite{FrachebourgKrapivsky96,BenNaim-etal96}. 

We have already explained how the asymptotic behaviour of the space-time dependent 
correlation functions can be obtained in a way that exploits the special properties of the master equation. 
An alternative treatment of the many-body correlation functions uses an
equivalence between the voter model  
and an auxiliary process of annihilating random 
walks~\cite{CoxDurrett95,CoxGriffeaths86,ScheucherSpohn88}.  
By using this approach, Scheucher and Spohn
obtained  the same result for the dynamical structure factor in the small $k$ and long-time limits
for $d$=2
\begin{equation}
S(\boldsymbol{k},t) \sim  \frac{1}{\ln(  t/\tau )} \frac{ 1}{\kappa^2(   t / \tau ) } 
\ F \left( \frac{ |\boldsymbol{k}| }{\kappa(  t/ \tau )} \right)
\label{SF_scaling_form}
\end{equation}
with $\kappa(t) \propto t^{-\frac{1}{2}}$ 
and  $F(u)=\frac{1}{2u^2}(1-\mathrm{e}^{-u^2})$, as found by employing the master equation formalism.
From here one recovers the asymptotic form for $C(\boldsymbol{x},t)$ in Eq.~(\ref{correlf_asympt3}).

\section{Numerical analysis}
\label{sec:numerical}

In this Section we present our numerical results. We first compare them to the analytical ones 
recalled in Sec.~\ref{sec:analytic} for infinite size systems and we later 
focus our attention on finite size effects. 

We define the model on a square lattice with linear size $L$ and 
periodic boundary conditions. In all cases we start the dynamics at time $t=0$ with a random initial 
condition with $s_{\boldsymbol x} = \pm 1$  with probability a half. 

One unit of time ({\it i.e.} $ \tau $) corresponds to $ L^2$ 
spin-flip attempts. As the system coarsens the number of flippable spins decreases and more attempts are necessary to change
 the configuration significantly. In order to accelerate the simulations we used a continuous time Monte Carlo algorithm with the 
 voter model dynamic rule.
Unless otherwise stated, the quantities that we present below were averaged over $ 10^5$ samples.

As we will be particularly concerned with the geometric properties of the coarsening process, let us give here a number 
of definitions that we will use in the rest of this Section.
We define a cluster or geometric domain as the ensemble of first-neighbour parallel spins. The cluster area
is the number of spins belonging to it. Any such domain is surrounded by an interface that corresponds to 
the ensemble of broken first-neighbour links surrounding the 
cluster. The total interface length (external plus internal) is the number of such oppositely oriented spin pairs. 

\subsection{Snapshots}
\label{subsec:snapshots}

\begin{figure}[h]
        \centering

      	\subfloat[$ t=4$]{%
                \includegraphics[width=3.5cm]{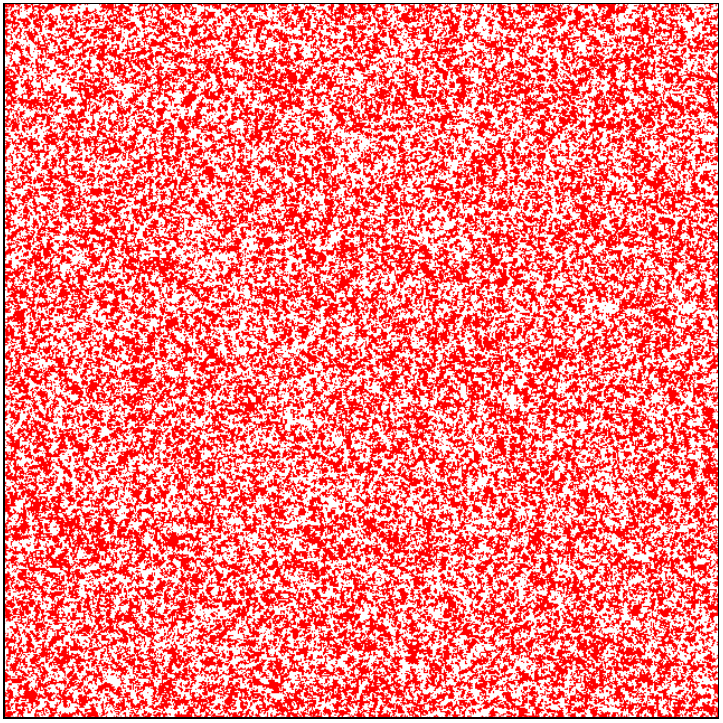}
	        \label{A}
	}%
      	\subfloat[$t=64$]{%
                \includegraphics[width=3.5cm]{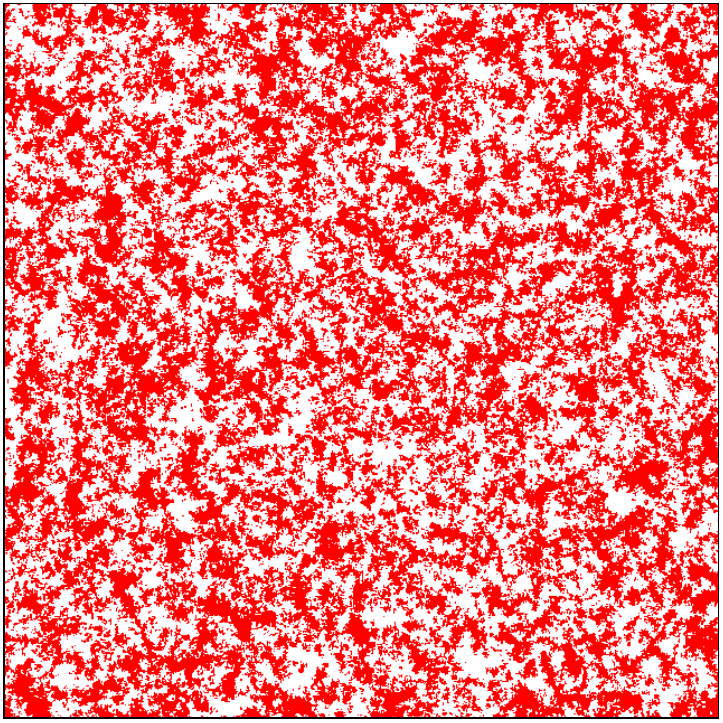}
	        \label{B}
      	}%
      	\subfloat[$t=512$]{%
                \includegraphics[width=3.5cm]{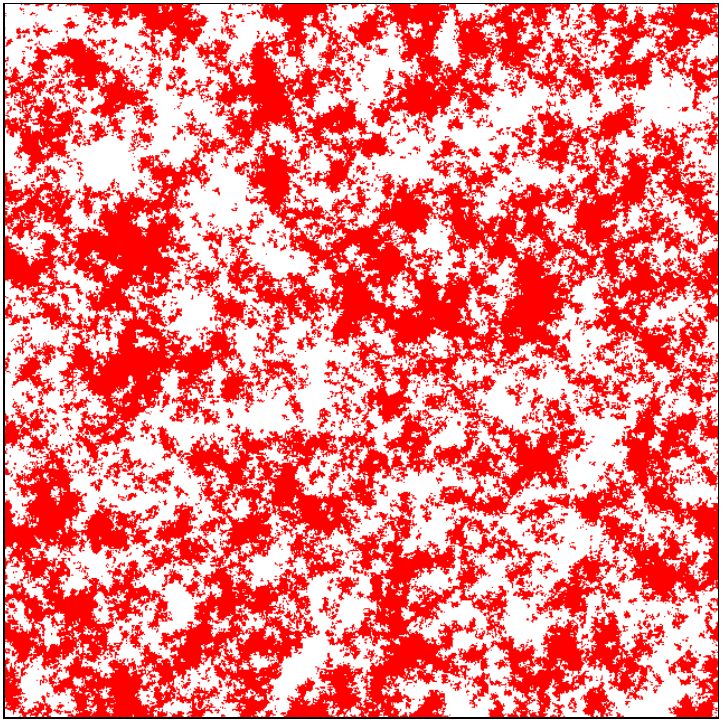}
	        \label{C}
	}
	\subfloat[$t=4096$]{%
                \includegraphics[width=3.5cm]{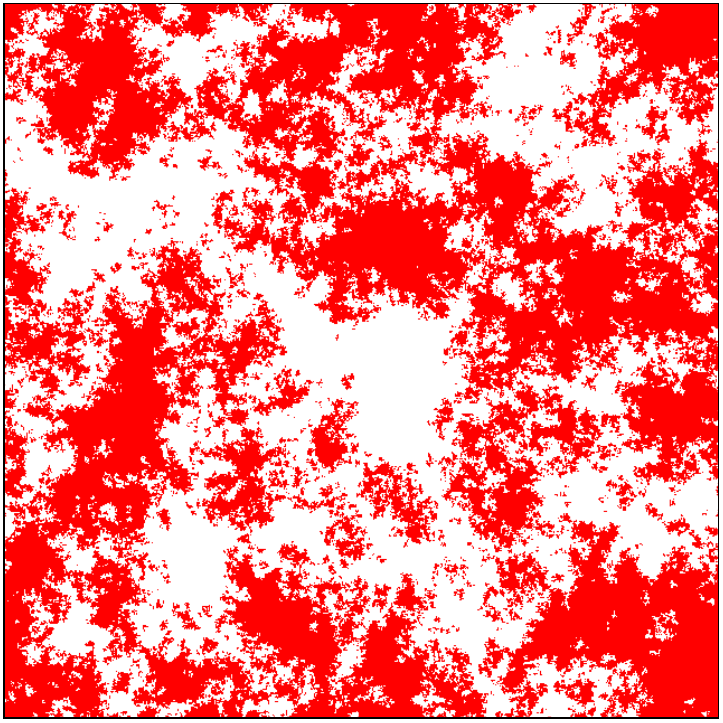}
	        \label{D}
	}%

	\subfloat[$ t=4$]{%
                \includegraphics[width=3.5cm]{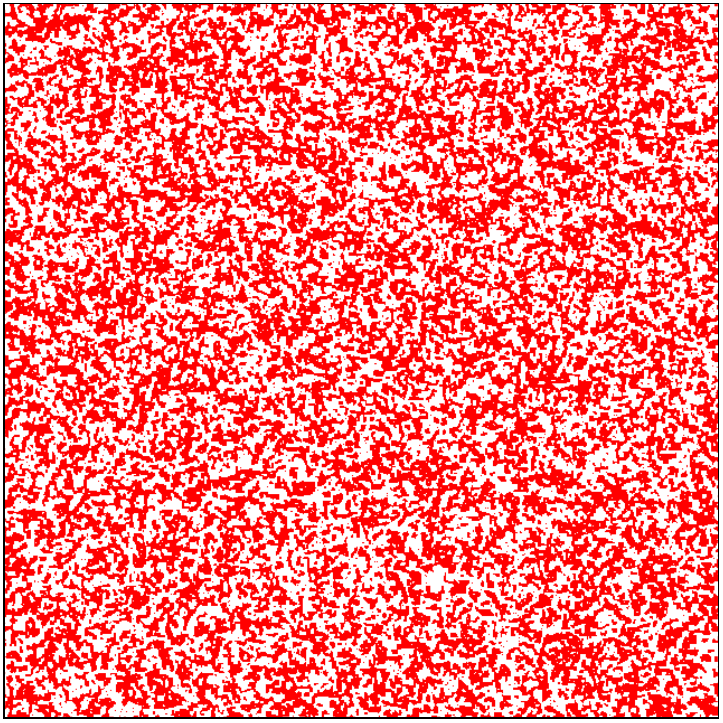}
	        \label{A}
	}%
	\subfloat[$t=64 $]{%
                \includegraphics[width=3.5cm]{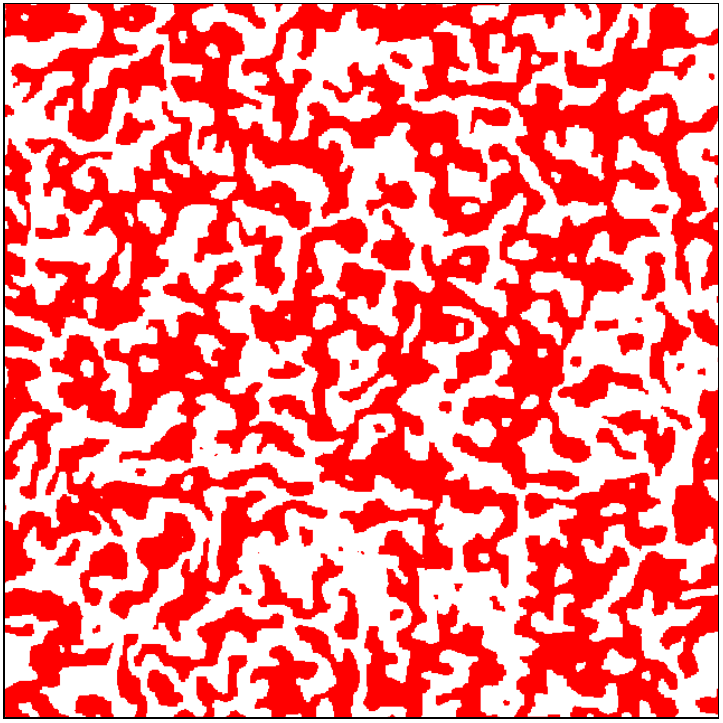}
	        \label{B}
	}%
	\subfloat[ $t=512$]{%
                \includegraphics[width=3.5cm]{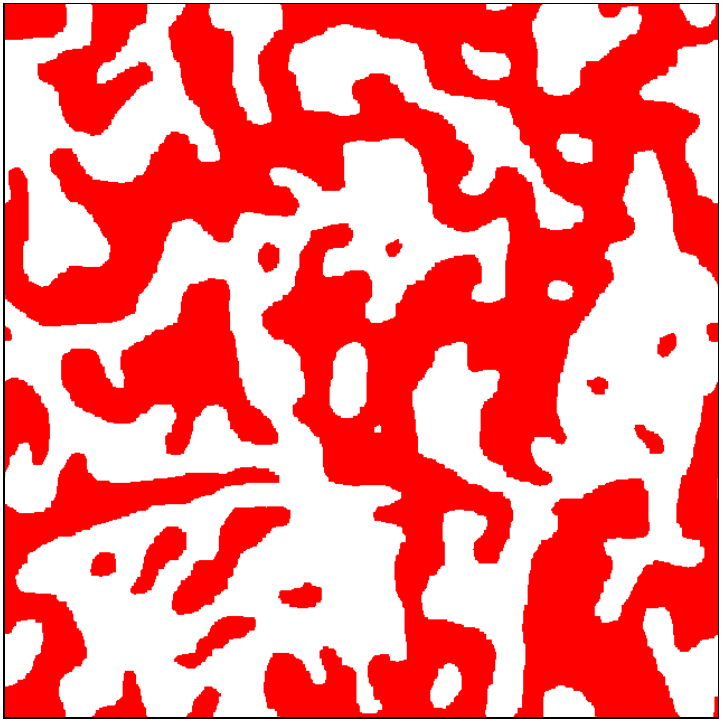}
	        \label{C}
	}%
      	\subfloat[ $t=4096$]{%
                \includegraphics[width=3.5cm]{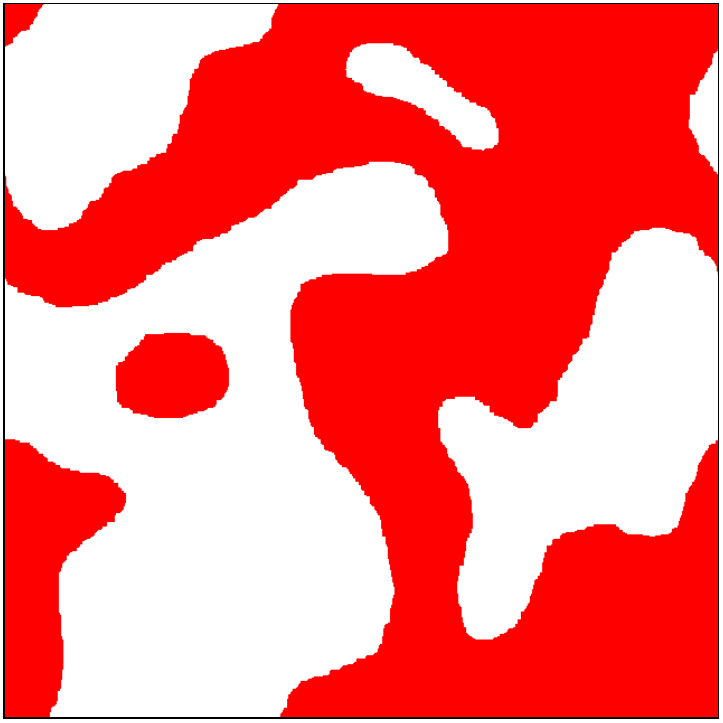}
	        \label{D}
	}%

	\subfloat[$ t=4$]{%
                \includegraphics[width=3.5cm]{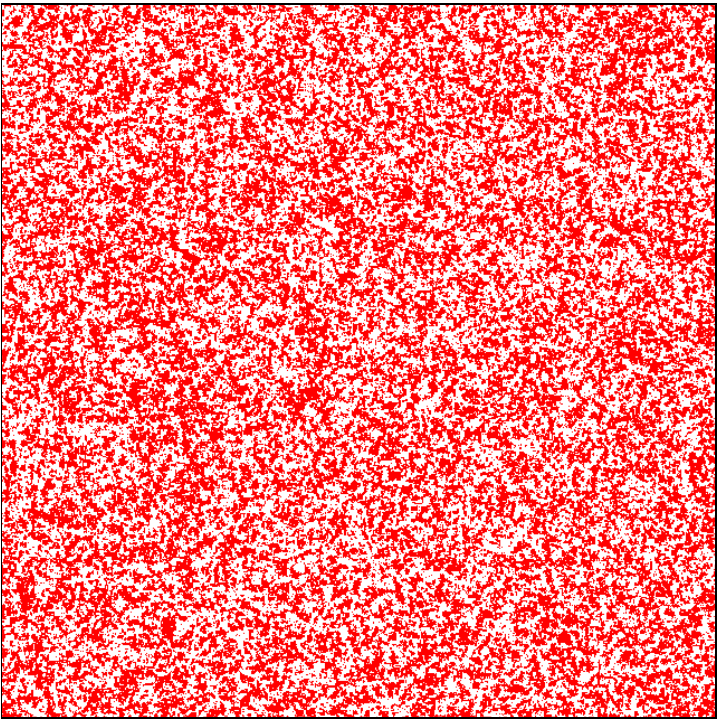}
	        \label{A}
	}%
	\subfloat[$t=64 $]{%
                \includegraphics[width=3.5cm]{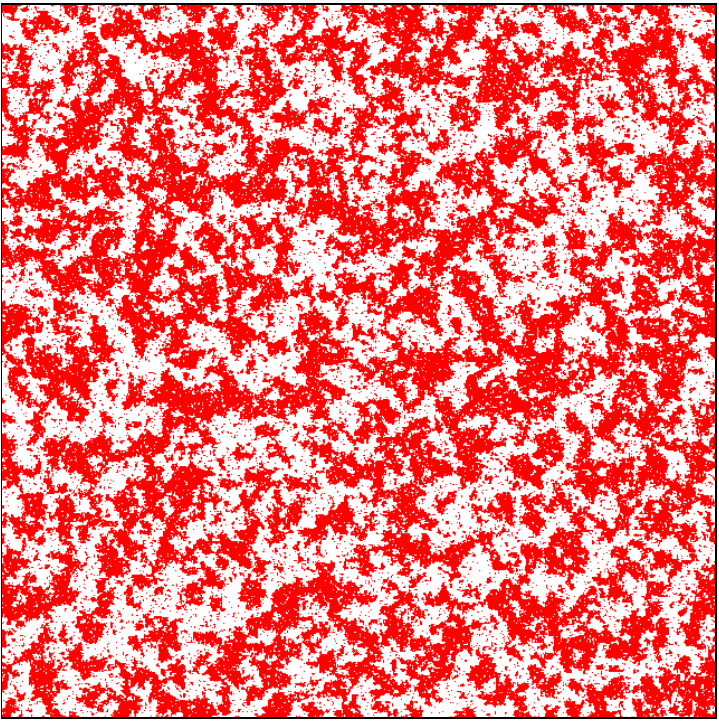}
	        \label{B}
	}%
	\subfloat[ $t=512$]{%
                \includegraphics[width=3.5cm]{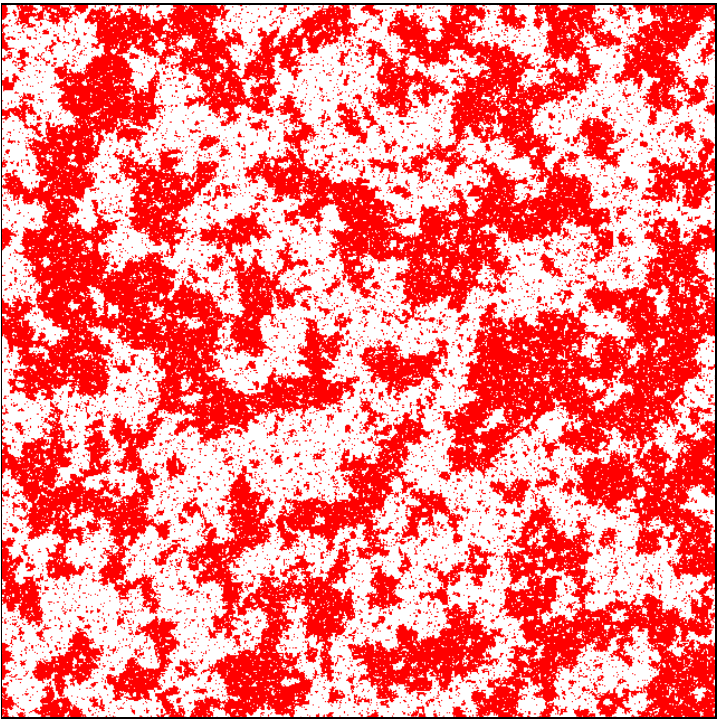}
	        \label{C}
	}%
      	\subfloat[ $t=4096$]{%
                \includegraphics[width=3.5cm]{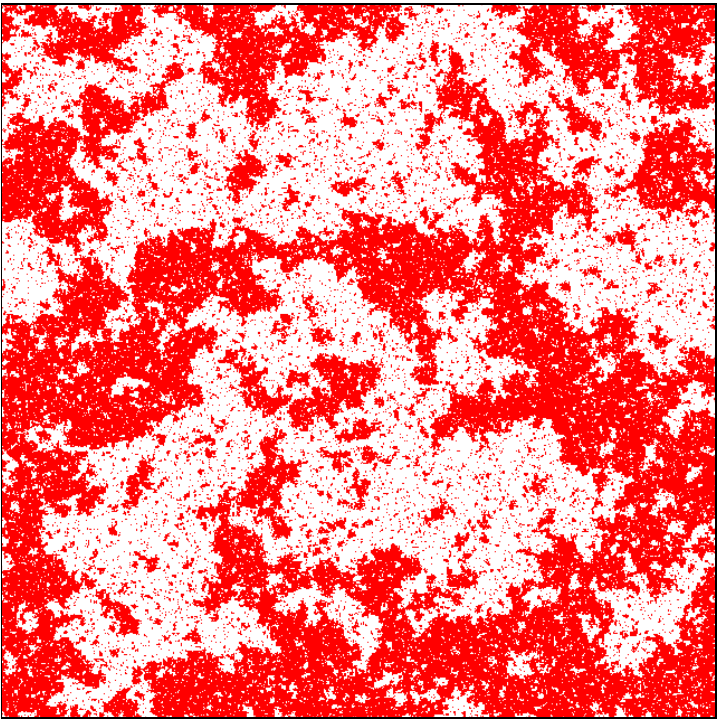}
	        \label{D}
	}%

        \caption{\footnotesize{Snapshots of the voter (first row) and Ising  (second and third rows) models
        on a $2d$ square lattice with linear size $L=640$ and periodic boundary conditions. 
        The Ising model has been quenched to $T=0$ (second row) and $T_c$ (third row). 
        The images were taken at the times indicated below the snapshots.
        }}
\label{fig:snapshots}
\end{figure}

In Fig.~\ref{fig:snapshots} we show three series of snapshots of the bidimensional voter model (first row) and the ferromagnetic 
Ising model at times $t=4, \ 64, \ 512, \ 4096$ (henceforth all numerical times are expressed in MCs and we omit this time unit 
to lighten the notation). The Ising model (IM) has been quenched to zero temperature (second row) and the critical point (third row) and it 
evolves with a heat-bath Monte Carlo algorithm. Red and white points represent the two spin configurations. The 
snapshots illustrate the coarsening phenomenon induced by the different microscopic dynamics.
In the case of the $2d$IM instantaneously quenched to $T=0$ the dynamics are purely curvature-driven: 
for sufficiently long time, all the interfaces move with a local velocity that is proportional to the local curvature~\cite{Bray94,Puri09}. 
As a result the interfaces tend to disappear independently of one another, {\it i.e.} there are no coalescence
processes. Instead, in the voter model the dynamics are driven by interfacial noise. In other words,
if the initial configuration consisted of a single flat interface between two domains of opposite opinion, 
opinions would slowly diffuse from one domain into the other and, after a sufficiently long time, the original sharp interface
would become a diffuse interface. As one can see from the snapshots, phase-ordering still occurs  but the resulting
domains are very jagged and preserve their fractal geometry even at the late stages of evolution.
Note, however, that the dynamics of the zero temperature Ising model and the voter model have one important feature in common,
namely, they are both characterised by the absence of bulk fluctuations. But they also show one important difference
in the morphological properties of their interfaces. 
Indeed, the domain walls in the voter model are more similar to the ones in the critically quenched Ising model,
shown in the third series of snapshots in the same figure, than to the ones in the Ising model evolving at any subcritical temperature.
The critical configurations are, though, plagued with bulk fluctuations, and these are   absent in the voter model.

\begin{figure}
        \centering
	\subfloat[$ t=4096 $]{%
                \includegraphics[width=3.5cm]{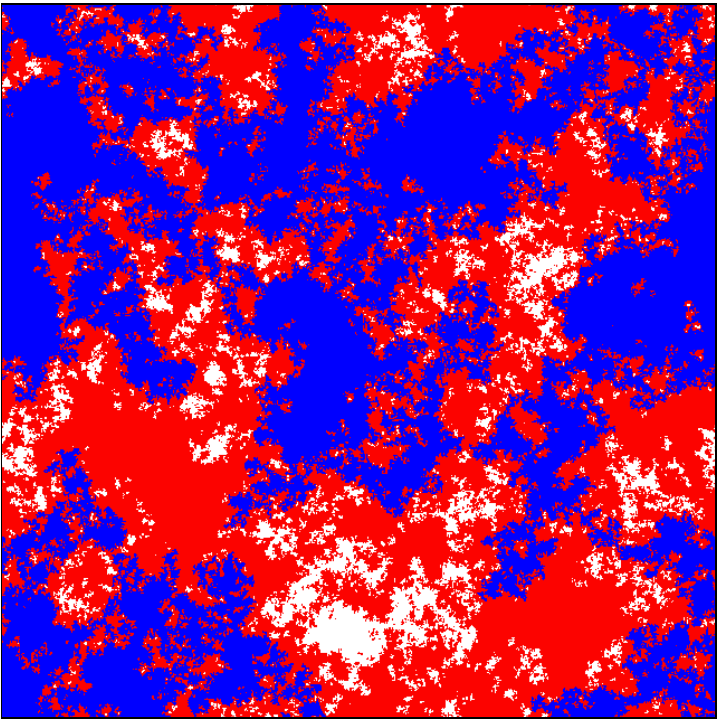}
	        \label{A}
	}%
	\subfloat[$ t=8192$]{%
                \includegraphics[width=3.5cm]{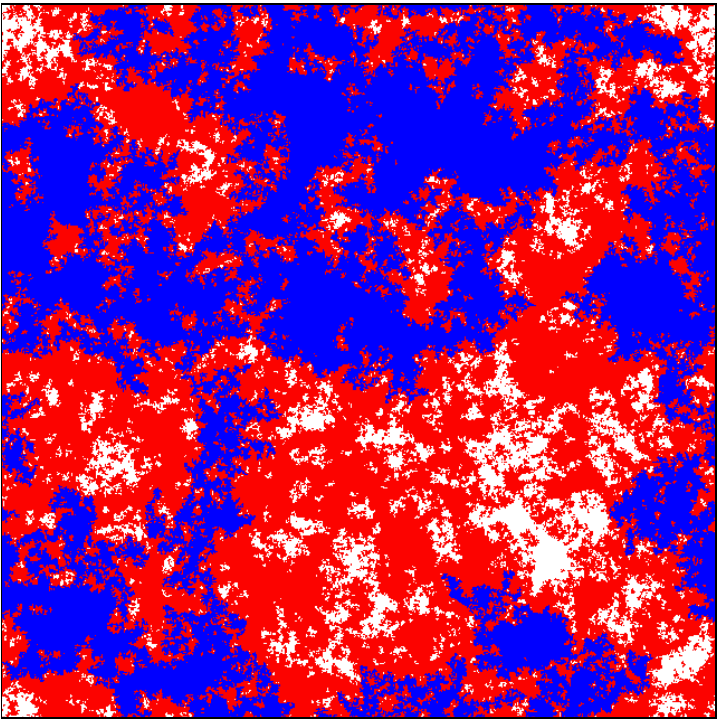}
	        \label{B}
	}%
	\subfloat[$ t=16384 $]{%
                \includegraphics[width=3.5cm]{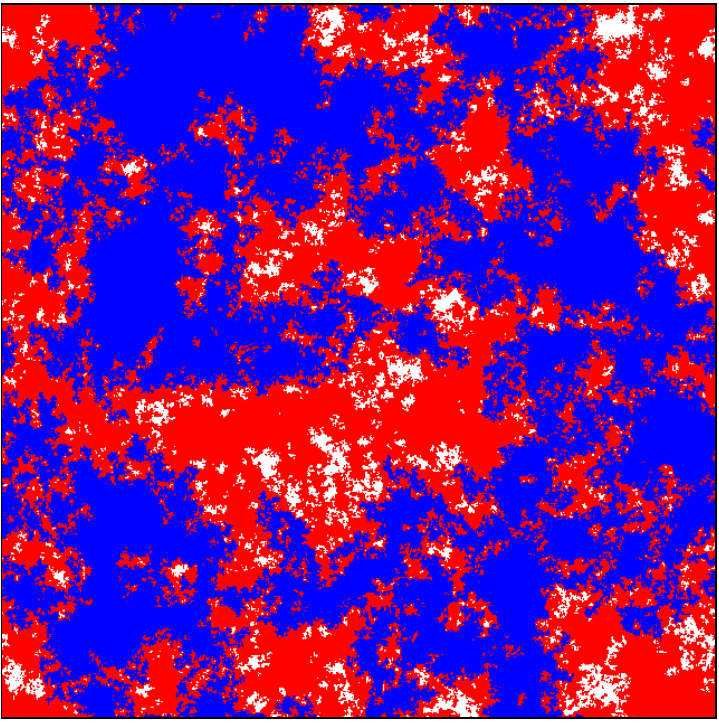}
	        \label{C}
	}%
      	\subfloat[$ t=32768 $]{%
                \includegraphics[width=3.5cm]{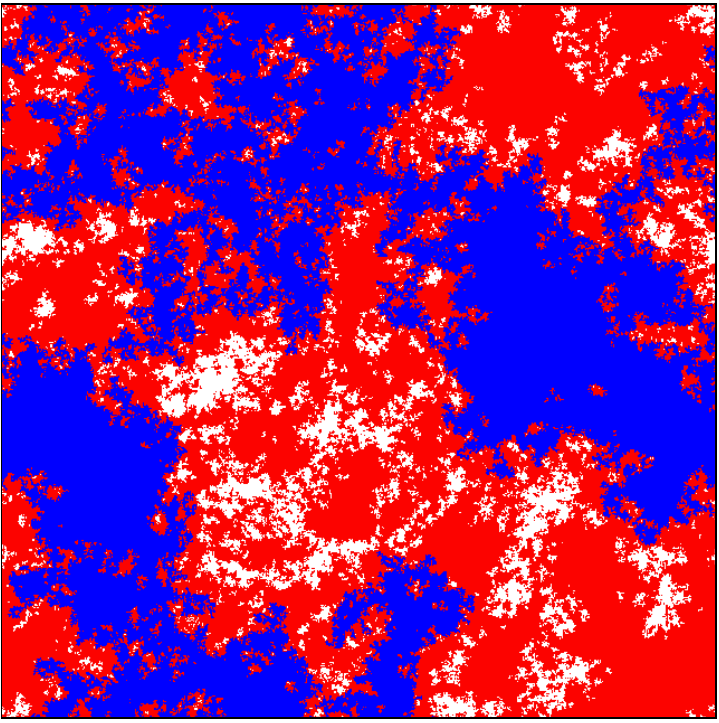}
	        \label{D}
	}%

      	\subfloat[$ t=65536 $]{%
                \includegraphics[width=3.5cm]{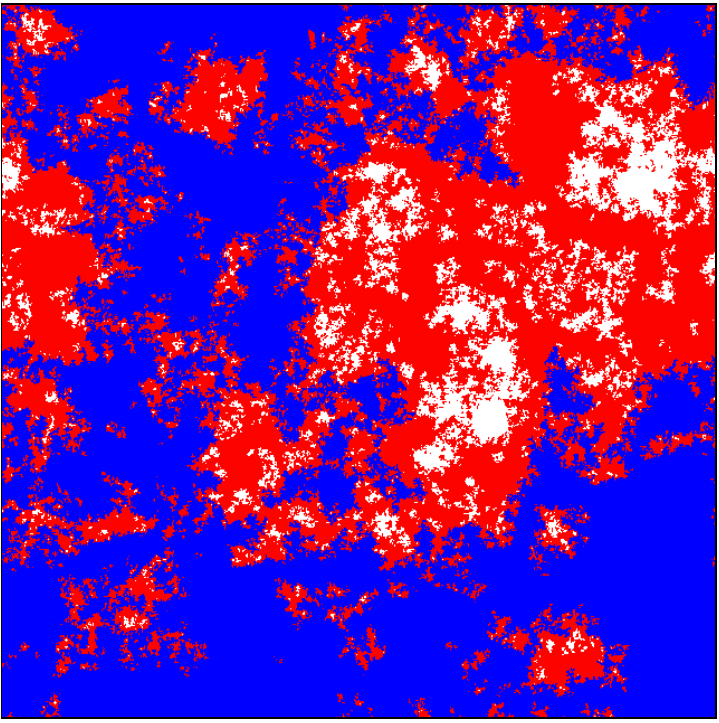}
	        \label{A}
	}%
      	\subfloat[$ t=131072 $]{%
                \includegraphics[width=3.5cm]{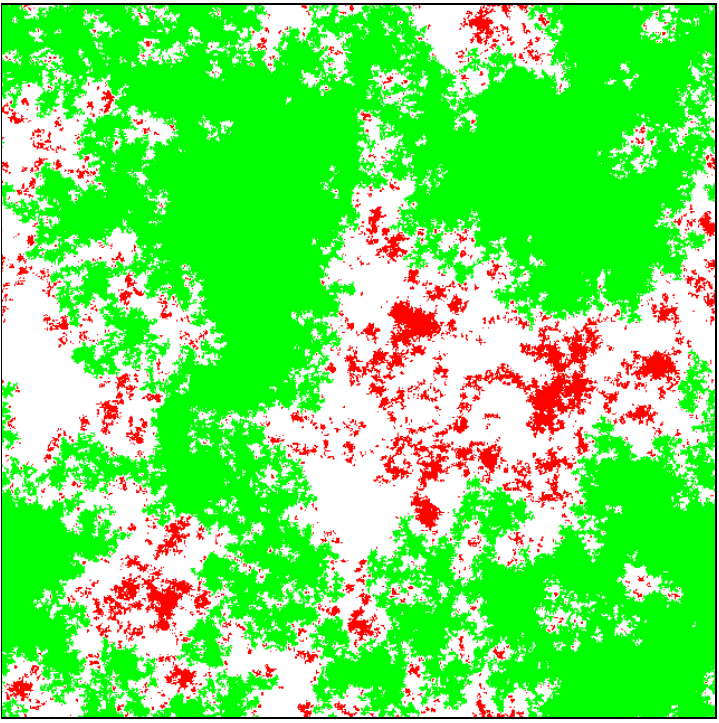}
	        \label{B}
      	}%
      	\subfloat[$ t= 262144 $]{%
                \includegraphics[width=3.5cm]{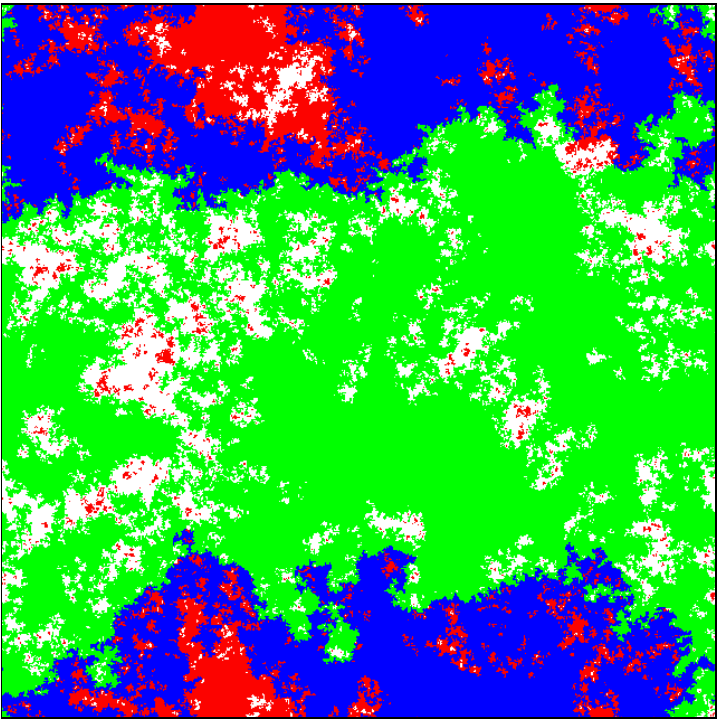}
	        \label{C}
	}
	\subfloat[$ t= 524288 $]{%
                \includegraphics[width=3.5cm]{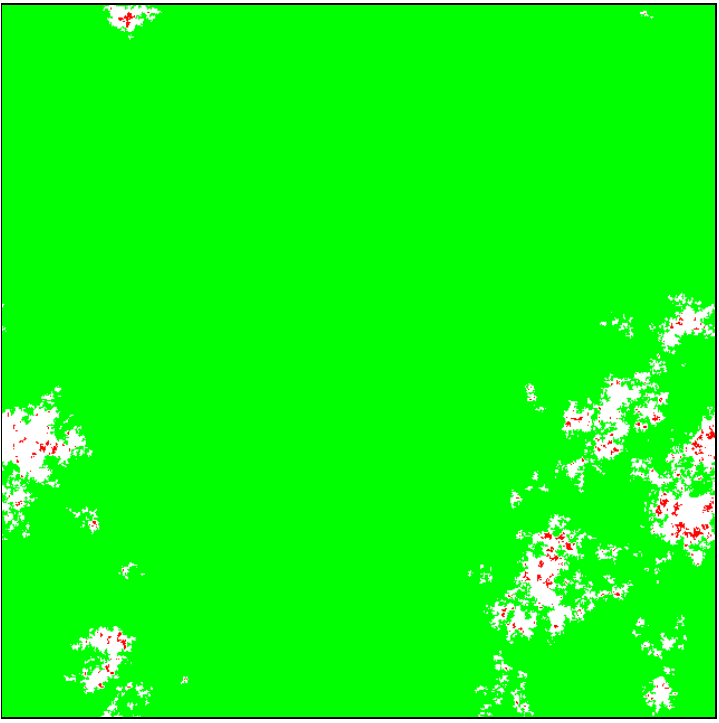}
	        \label{D}
	}%
        \caption{\footnotesize{Snapshots of the voter model with 
        linear size $L=640$ and periodic boundary conditions.
        The times shown are $t_i = 2^i $, with $ i= 12 - 19$. With the same convention as in Fig.~\ref{fig:snapshots}, the $ +1$ voters
        are coloured in red, while the $ -1$ in white. We highlight here the percolating clusters with different colours.
        Percolating clusters of $ +1$ ($-1$) opinion are green (blue). A single domain of ``$ +1$'' voters was reached at 
        $ t \approx 6.7 \cdot 10^5 $.}}
\label{fig:pc_snapshots}
\end{figure}

\par
In Fig.~\ref{fig:pc_snapshots}	 we display a series of snapshots of the voter model for even longer times than the ones 
used in Fig.~\ref{fig:snapshots} highlighting the percolating clusters of the two types. These configurations could be 
compared to the ones shown in~\cite{BlCoCuPi14} for the $2d$IM quenched to $T=0$. We note that the identity and 
form of the percolating clusters are not preserved in the first seven snapshots, until the system enters 
the late stage of evolution and finds full consensus.

\par
In Sec.~\ref{subsec:largest-cluster} we 
identify the largest cluster in the system and we study several of its properties letting us obtain in this way the 
exponent linking the system size to the time needed to reach percolation, and the fractal dimension of the 
percolating cluster and the one of its  perimeter. 

\par
In the case of a finite lattice with periodic boundary conditions
one can distinguish two types of domains: the ones that are homotopic to a point 
on the torus and the ones that wrap around the hole 
and cannot be completely shrunk without breaking into disconnected pieces. 
Even though the former can have a linear size comparable or even bigger than the one of the system, 
we identify the percolating clusters as the ones that wrap around the torus hole only.

\subsection{Interface density}
\label{subsec:interface_density}

In Sec. \ref{sec:analytic} we provided an expression for the long-time behaviour of the fraction of active interfaces.
In the disordered initial condition $\rho(0) \simeq 1/2$. 
In $d=1$ the voter model is equivalent to the Glauber IM and $\rho(t)$ decays as  $t^{-1/2}$ while in $d>2$ the 
density of interfaces converges to a constant, $ \rho(t) \sim a-b t^{-d/2}$. In $d=3$ the model has, then, 
blocked configurations asymptotically, as in the
$2d$ and $3d$ IM at $T=0$~\cite{OlKrRe11a,OlKrRe11b}.
In the case of $d=2 $, $ \rho(t) $ decays logarithmically and several authors~\cite{Meakin-Scalapino87,Evans-Ray} tried to
study  this particular behaviour with Monte Carlo simulations.
By starting from Eq.~(\ref{rho_eq}) it is possible to obtain a more refined estimate of $ \rho(t)$~\cite{FrachebourgKrapivsky96}, 
\begin{equation}
\rho(t) = \frac{\pi}{2\ln t + \ln 256} + O\left( \frac{\ln t}{t} \right)
\; .
\label{eq:rho-analytical}
\end{equation}
This result
has to be contrasted to the algebraic decay, $\rho(t) \sim t^{-d/z_d}$, of curvature driven domain growth.
For instance, in the $2d$IM model this 
same quantity decays as $\rho(t) \simeq \ell(t)^{-d} = t^{-d/z_d} = t^{-1}$, with $\ell(t)$ the characteristic growing length 
and $z_d=2$.

\begin{figure}
\centering
\centering
\includegraphics[scale=0.8]{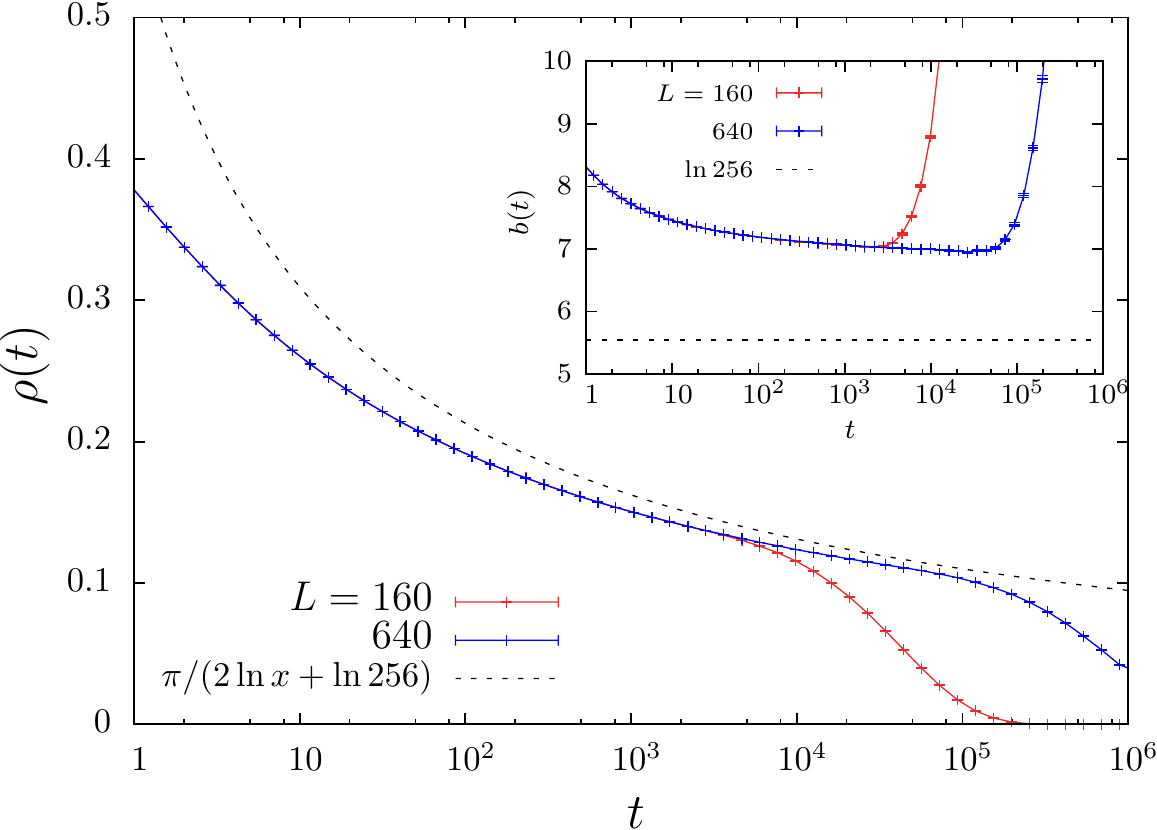}
\caption{\footnotesize{Concentration of active interfaces $\rho(t) = (1-\langle s_{\boldsymbol{x}}(t) s_{\boldsymbol{x} + \boldsymbol{\mathrm{e}}_i}(t) 
  \rangle)/2$ as a function of time. 
Data for $L=160$ and $640$ are presented in a  log-linear form. The error-bars are smaller than the symbol sizes. The dotted (black) line 
represents the analytic asymptotic form $\rho(t) = a/(2 \ln t + b)$ with $a=3.14$ and $b=5.54$. 
  In the inset we show $b = \pi/ \rho(t) - 2 \ln t$ as a function of $t$ for 
 different lattice sizes together with the analytic value $5.54$ shown with a horizontal dashed line. See the main text for a discussion.
 }
}
\label{fig:av_RHO}
\end{figure}

\begin{figure}
\centering
\subfloat[]{
\includegraphics[scale=0.6]{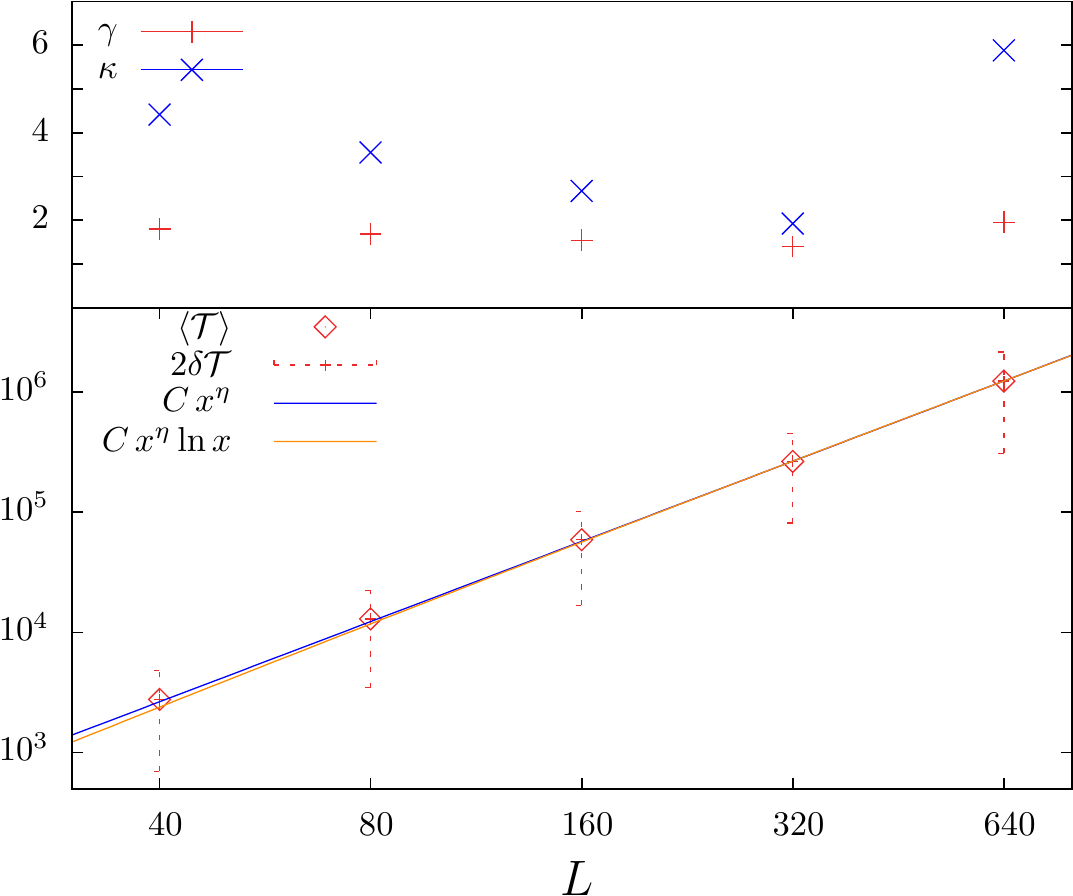}
}%
\subfloat[]{
\includegraphics[scale=0.67]{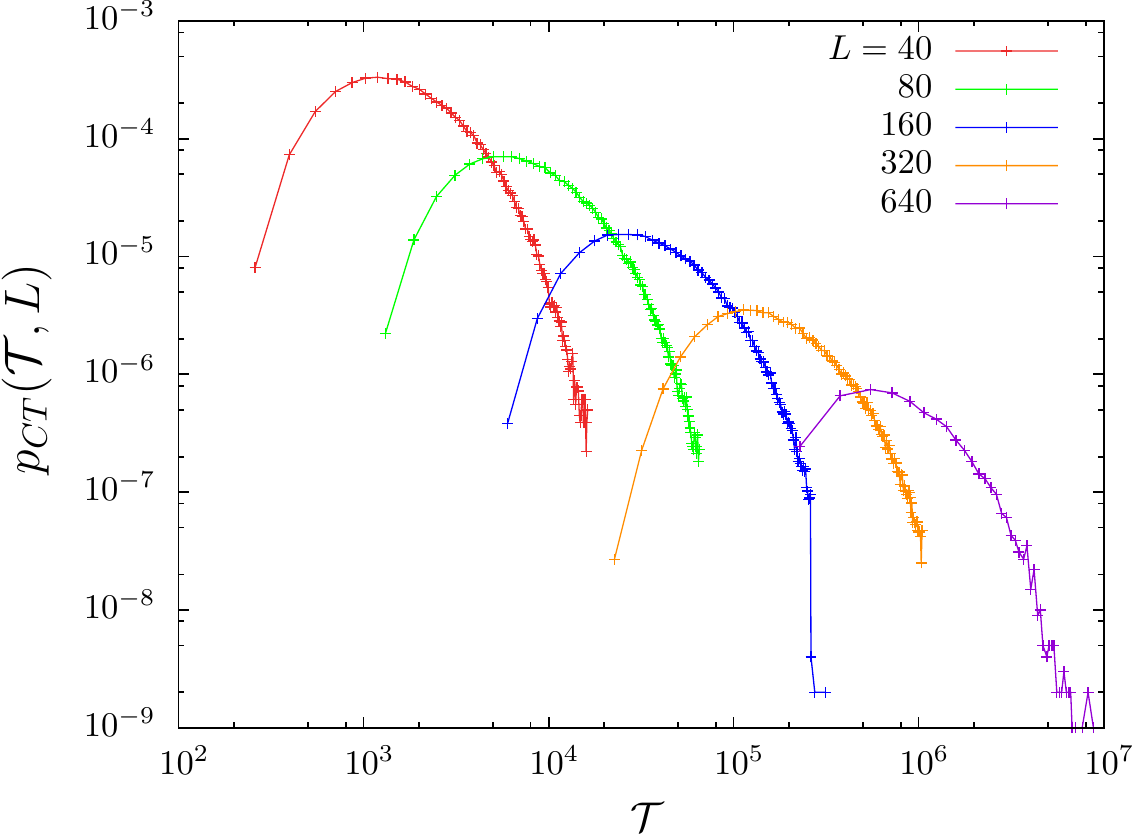}
}%
\caption{\footnotesize{(a) The average consensus time $ \langle \mathcal{T} \rangle $ (red diamonds) for different values of the lattice linear size $ L $ in a log-log plot.
The value of $\langle \mathcal{T} \rangle$ has been computed over approximately $ 10^5$ samples for the smaller sizes, and over about $7 \cdot 10^3$ samples
for $ L=640 $. We report also the approximate width of the distribution,
computed as the standard deviation of the data collected, $ \delta \mathcal{T} =  
[N^{-1}_s  \ \sum^{N_s}_{i=1} \left( t_i - \langle \mathcal{T} \rangle \right)^2 ]^{1/2}$, as vertical dashes with length $2\delta{\cal T}$ 
on each data point.
Roughly, $ \delta \mathcal{T} / \langle \mathcal{T} \rangle \simeq 0.74 $ for all the sizes. The function $ C \, x^{\eta} $ was fitted to the
few data points available yielding $\eta \simeq 2.21 $ and $ C \simeq 0.75 $. 
By introducing a logarithmic prefactor, $ C \, x^{\eta} \ln x$, the fit yields instead $\eta \simeq 2.05 $ and $ C \simeq 0.33 $.
In the upper part of the same panel we report the skewness $\gamma$ and the kurtosis $\kappa$ of the sampled distributions
in order to quantify the deviations from a Gaussian form. (b) The distribution of consensus times for different system sizes $L$ given in the key.
}}
\label{fig:av_CT}
\end{figure}

In Fig.~\ref{fig:av_RHO} we present numerical data for $\rho(t)$ in a voter model with linear size $L=160$ and $640$ with times
reaching $t =10^6$. The analytic result in the asymptotic limit  $ f(t)=a/(2 \ln t + b)$ with $a=\pi $ and $b=\ln 256 \simeq 5.54$
accompanies the data as a dotted (black) line. 
We have performed detailed fits of the data finding that the parameter $a$ approaches the analytic 
value quickly. We then fixed $a=\pi$ and we measured the parameter $b$ by studying
$b=\pi/\rho(t) - 2 \ln t$ as a function of $t$ for different system sizes ($\rho(t)$ are the measured values). We show the
result of this analysis in the inset to the same figure. The approach to the analytic value shown as a black dotted horizontal line 
is indeed very slow. This fact explains why several authors did not see the asymptotic law in their numerical 
data and used instead a different logarithmic form, $ \rho(t) = C \ln^{-\sigma} t$, with an 
 effective exponent $ \sigma \approx 0.6$, to fit their data~\cite{FrachebourgKrapivsky96,Evans-Ray,Evans}.

\subsection{Consensus time}
\label{subsec:consensus-time}

\par
With much longer simulations we were able to measure the average consensus time,  $\langle \mathcal{T} \rangle = N_s^{-1} \sum_{i=1}^{N_s} \mathcal{T}_i$
with $\mathcal{T}_i$ the time required by the $i$-th sample
to reach full alignment and $ N_s$ the total number of samples.
In Fig.~\ref{fig:av_CT}~(a) we show the results obtained by averaging over at least $ 7 \cdot 10^3 $
realisations of the dynamics for each value of the lattice size $ L$.
The averaged consensus time  approximately follows the law $ \langle \mathcal{T} \rangle (L) \sim L^2 $. 
However, in Sec.~\ref{sec:analytic} we recalled that the correlation functions suffer from logarithmic correction. 
Therefore, we tried to take into account this kind of correction
by fitting   the function $C \, L^{\eta} \ln L$ to the data $(L,\langle \mathcal{T}\rangle (L))$ and 
we obtained $ \eta = 2.05 \pm 0.01 $, $ C = 0.33 \pm 0.02$, and a better agreement with the numerical data
than with the pure power law. 

\par	
An estimate of the characteristic width of the probability distribution of the consensus time is given by
the standard deviation, $ \delta \mathcal{T} =  
[  N^{-1}_s  \ \sum^{N_s}_{i=1} \left( \mathcal{T}_i - \langle \mathcal{T} \rangle \right)^2 ]^{1/2}$. 
The relative standard deviation $ \delta \mathcal{T}/\langle \mathcal{T} \rangle $ was found to lie
in the interval $ 0.70 - 0.75 $ for every $ L $ and no particular dependence on the number of samples was observed
for $ N_s > 10^3$. To highlight this behaviour we added vertical dashes of width $2 \delta \mathcal{T}(L)  $ centred on each one of the data points  
in Fig.~\ref{fig:av_CT}~(a). We stress that these dashes do not represent any type of error on the numerical value of the average
consensus time, but only a measure of the average dispersion of our data on the available population of samples.

\vspace{0.5cm}

 \begin{figure}[h]
\centering
\includegraphics[scale=0.65]{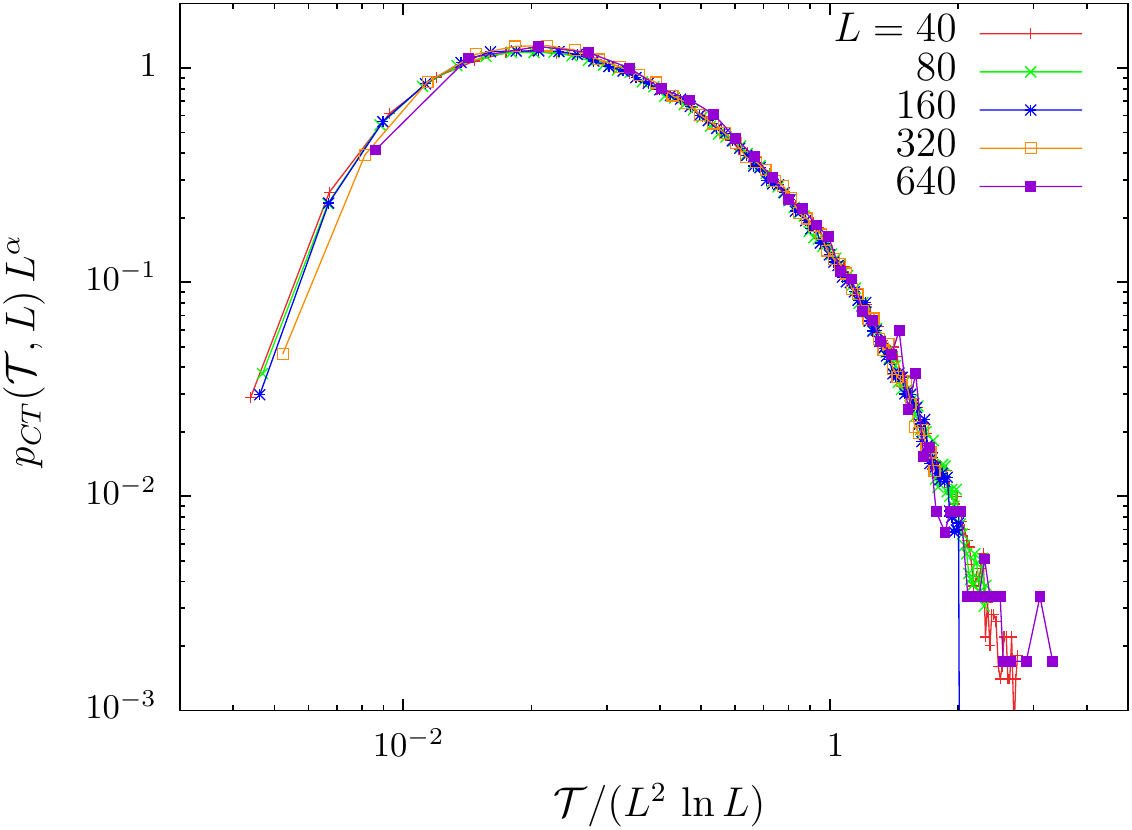}
\caption{\footnotesize{Rescaled histogram of the consensus time $ p_{CT} (\mathcal{T},L)$ for different lattice sizes. Time $\mathcal{T}$ is rescaled by
$ L^2 \, \ln{L} $, while the value of $ p_{CT} (\mathcal{T},L)$ is multiplied by $L^{\alpha}$. 
The best collapse of the different curves was obtained for $\alpha \simeq 2.22$. 
}}
\label{fig:rescaled_CT_histo}
\end{figure}

In Fig.~\ref{fig:av_CT}~(b) we show the histogram of consensus times $ p_{CT} (\mathcal{T},L)$ for the different lattice sizes that have been simulated.
The curves have approximately all the same shape when plotted in a log-log scale, so it is reasonable to 
assume the following scaling {\it Ansatz}
\begin{equation}
 p_{CT} (\mathcal{T},L) = L^{-\alpha} \, \mathcal{P} \left( \frac{\mathcal{T}}{L^\beta \, \ln{L}}\right)
 \label{eq:ctime_scaling}
\end{equation}
 with exponents $\alpha$ and $\beta$ to be determined. Equation~(\ref{eq:ctime_scaling}) simply states that the probability distributions
 of the consensus time for finite systems with different size have identical form, apart from a prefactor, if time is rescaled
 by a typical time $ \mathcal{T}_{typ} \sim L^\beta \, \ln{L} $. This is a natural choice since 
 $  \langle \mathcal{T} \rangle \sim L^2 \, \ln{L} $ and thus we expect $\beta$ to be very close to 2.
 In Fig.~\ref{fig:rescaled_CT_histo} we show the result of the scaling for $\beta=2$ and $\alpha = 2.22$. 
 We could not appreciate significant improvements of the collapse for values of $\beta$ slightly different from $2$,
 so our assumption was confirmed. 

The upper part of Fig.~\ref{fig:av_CT}~(a) displays the skewness, 
$\gamma \equiv
N_S^{-1} \sum_{i=1}^{N_s} ( {\mathcal T}_i - \langle {\mathcal T} \rangle )^3 /  \delta{\mathcal T}^{3} $,
 and kurtosis,
 $\kappa \equiv 
N_S^{-1} \sum_{i=1}^{N_s} ( {\mathcal T}_i -  \langle {\mathcal T} \rangle )^4 /  
\delta{\mathcal T}^{4} -3$, 
  of $p_{CT}(\mathcal{T},L)$ as a function of system size. 
  The deviation from zero of the former 
  quantifies the asymmetry of $p_{CT}$ and of the latter gives a further  idea of the non-Gaussian
  character
  of the distribution. The skewness seems to have converged to a system size independent value
  that is slightly higher than one, while the kurtosis is still varying significantly for different system sizes.
The last data points, for $L=640$, are clearly not converged and many more samples 
would be needed to reach a good estimate for them.

\subsection{Persistence and autocorrelation}
\label{subsec:persistence}

\par
In general one defines $ P_n(t)$ as
the probability distribution for the number of opinion changes $ n $ experienced by a voter during the time interval $ ( 0, t)$, with 
$ n \in \{0,1,2,...\} $~\cite{BenNaim-etal96}. 
The first of these quantities, $ P_0(t)$, is equal to the fraction of voters who did not change opinion up to time $ t $, 
\vspace{0.25cm}
\begin{figure}[h]
\centering
	 \subfloat[]
	 {
	 \centering
\includegraphics[scale=0.6]{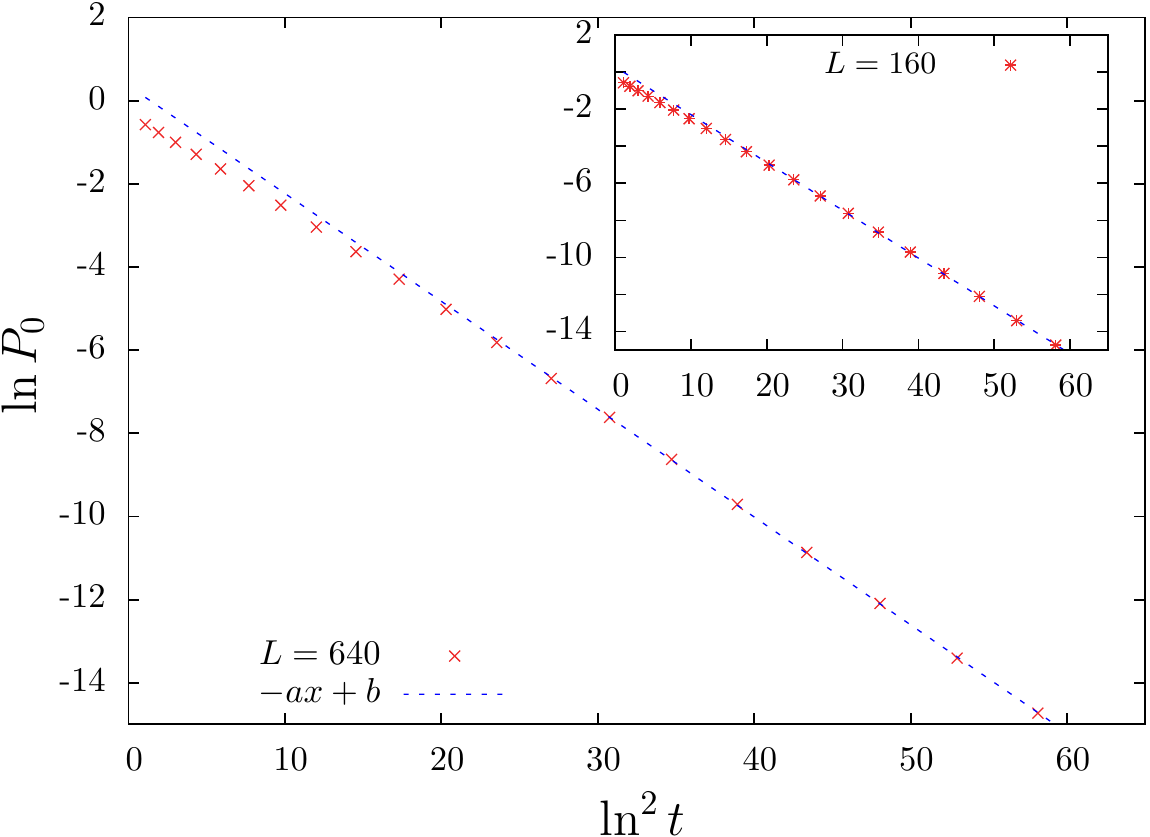}
}
	 \subfloat[]
	 {
	 \centering
\includegraphics[scale=0.6]{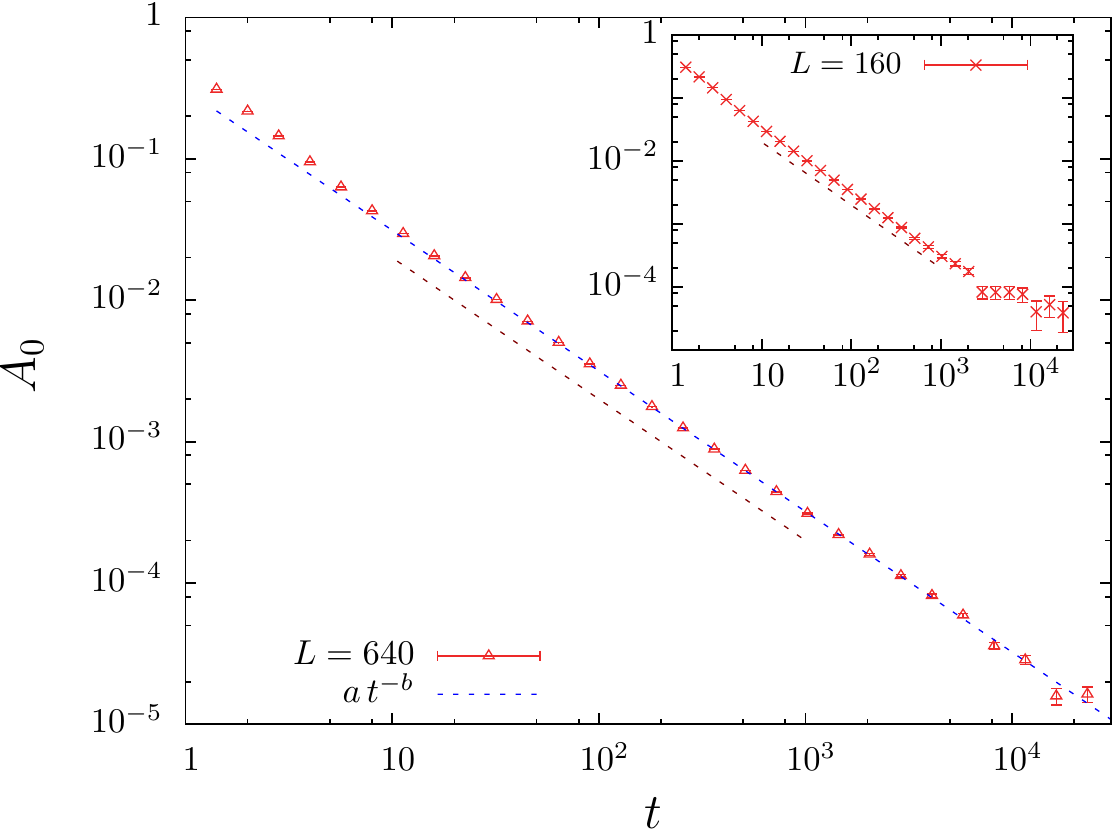}
}
\caption{\footnotesize{(a)~The logarithm of the persistence probability, 
  $ \ln P_0$, against $ \ln^2{  t } $, for a system of size $ L=640 $. A  linear fit to the data in this scale,  
   $ f(x) = b - a x $, gives $ a \simeq  0.26 $ and $ b \simeq 0.36  $ (the fitting curve is showed as a dashed blue line). 
   The persistence has been calculated as an average over
   $ 5 \cdot 10^4$ realisations of the dynamics. Data for $ L=160$ are shown in the inset along with their fitting curve. 
(b)~Time-dependence of the autocorrelation with the initial configuration, $ A_0$, in a double logarithmic scale. 
The function $ f(t) = a t^{-b} $  has been fitted to the numerical 
data, yielding $b = 0.995 $ and $ a = 0.308 $, which are both in agreement with what theory predicts, {\it i.e.}  
$ A(t) \sim \left( \pi t \right)^{-1} $, shown with a dark dashed segment.
In the inset we show data for $ L=160$ analysed in a similar fashion.
}}
\label{fig:av_A0}
\end{figure}
{\it i.e.}~the 
persistence probability~\cite{BrayMajumdarSchehr13}.
In terms of spins, it measures the fraction of sites that have not experienced any spin-flip up to time $ t $. 
In most statistical physics models~\cite{BrayMajumdarSchehr13}, 
the persistence decays in time with a power law $ t^{- \theta }$ with a new independent persistence exponent $\theta$. 
In the two-dimensional Glauber-Ising model
at zero temperature, the exponent $\theta$ has been evaluated numerically with high precision and it takes the 
value $ \theta \approx 0.199(2)$ for initial conditions with short-range correlations~\cite{BlCuPi14}.
The asymptotic behaviour of $ P_0 $ in the voter model in $ d \ge 2 $ was first found numerically~\cite{BenNaim-etal96} and 
then computed analytically with a mapping onto a continuum reaction-diffusion process and the use of field theoretical 
tools~\cite{HowardGodreche98}. In $ d=2 $, 
\begin{equation}
 P_0(t) \sim k \cdot \mathrm{exp} \left[ -a  \ln^2 t +\mathcal{O} \left( \ln t \right) \right] 
 \label{persistence_eq}
\end{equation}
for our choice $\tau d =2$.
The difference in the behaviour of the persistence between the $2d$IM and the $2d$ voter model was investigated 
by Drouffe and Godr\`eche~\cite{DrouffeGodreche99} who introduced a class of stochastic processes on a $ 2d $ square lattice 
that interpolate between these two. They also confirmed the unusual time-dependence of the persistence decay
in the $2d$ voter model with numerical simulations.

\par
In this context, we tried to recover the theoretical prediction in Eq.~(\ref{persistence_eq}) with simulations 
of the voter model with different sizes. We present data for $L=160$ and $L=640$ in Fig.~\ref{fig:av_A0}~(a).
By fitting the simulation data to the function in Eq.~(\ref{persistence_eq})
we found $ a \approx 0.26 $ and $ k \approx \mathrm{e}^{0.36} \simeq 1.44  $.
The estimated value of $ a $ is quite close to the theoretical value predicted by Howard and Godr\`eche ~\cite{HowardGodreche98}, who found
$ a \simeq 1/4 $ with corrections of order $ 0.01$. 

\vspace{0.5cm}

\begin{figure}[h]
\centering
	 \subfloat[]
	{%
	        \centering
                \includegraphics[scale=0.625]{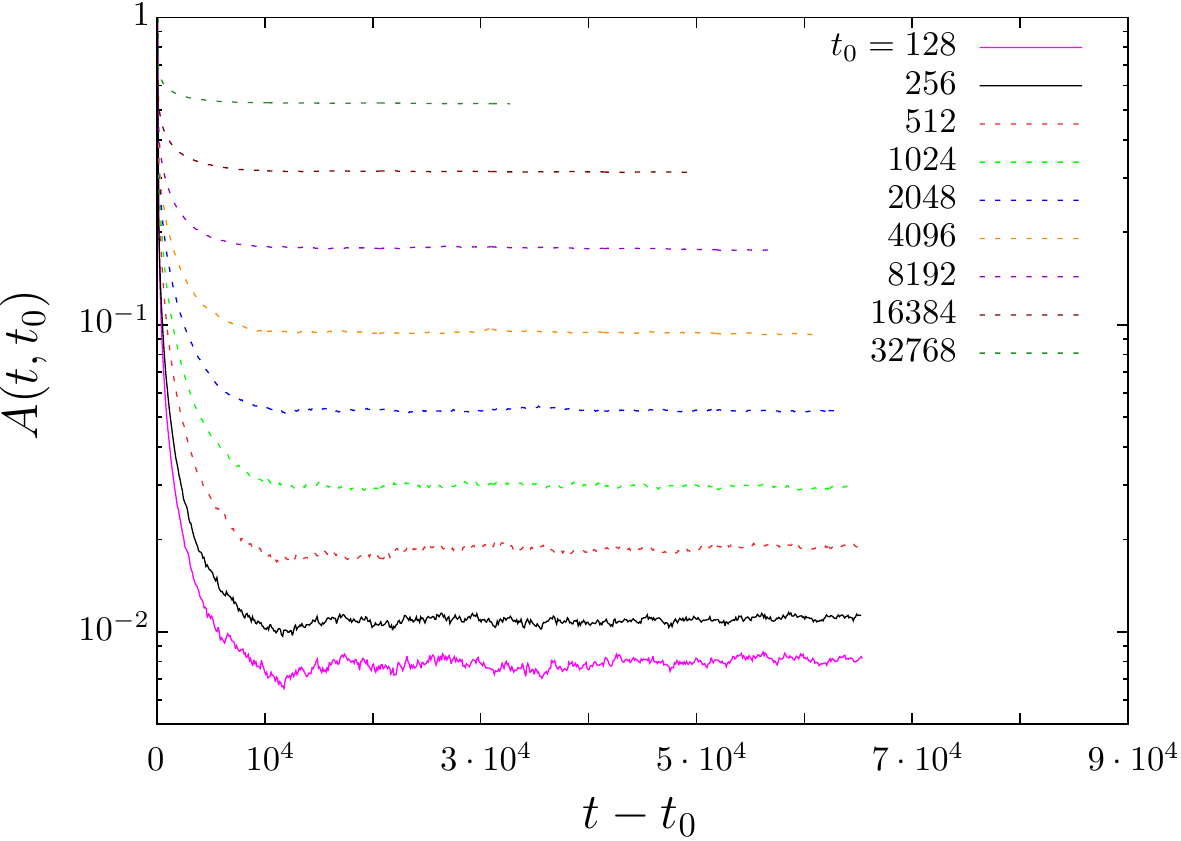}
	}%
	\quad
	 \subfloat[]
	{%
                \centering
                \includegraphics[scale=0.625]{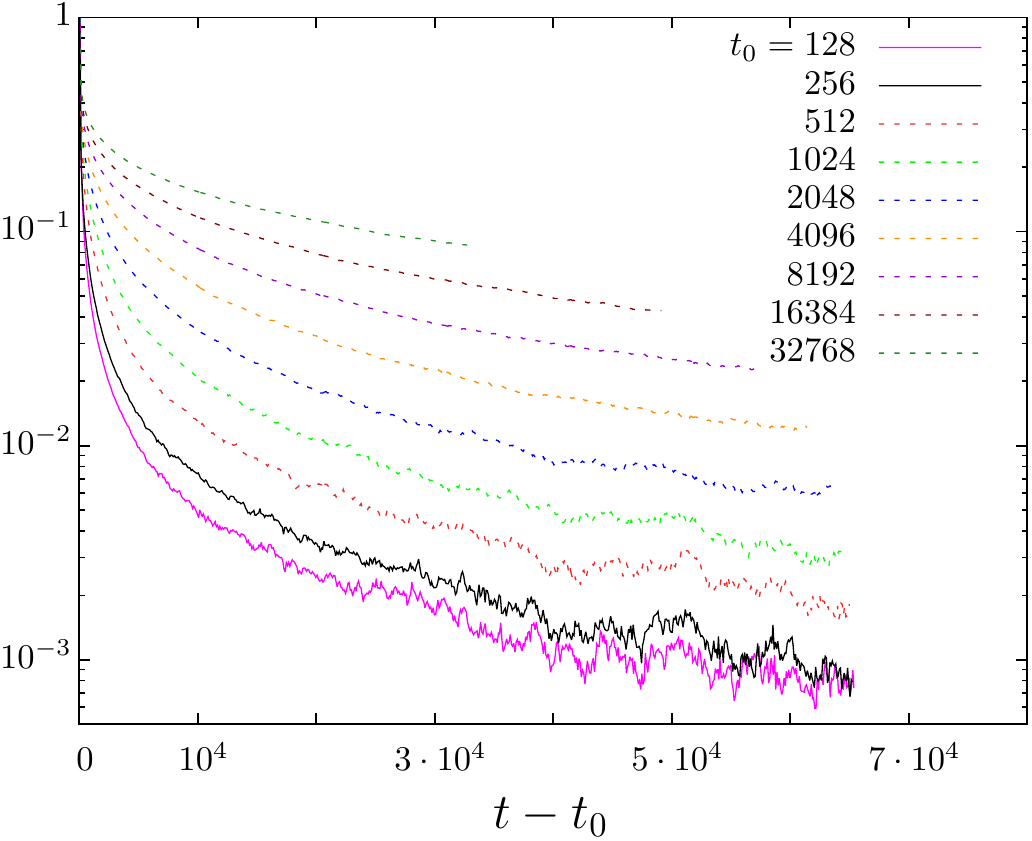}
	}%
        \caption{\footnotesize{The two-time autocorrelation function $ A(t,t_0) 
        $ against time-delay $t-t_0$ for different values of the waiting-time $ t_0 $. Panels (a) and (b) 
        show data for  systems with linear size $ L=160 $ and $L=640$, respectively. 
}}
\label{fig:autocorrelations}
\end{figure}%

\par
In Sec.~\ref{sec:analytic} we showed that the autocorrelation with a completely uncorrelated 
initial configuration, $ A_0 (t) $, has the asymptotic behaviour 
$ A_0(t) \simeq \left( \pi t \right)^{-1} $ in $ d=2 $, with the choice $ \tau = 1 $. 
In Fig.~\ref{fig:av_A0} we show numerical data for $A_0(t)$ in a system with linear size $ L=640 $. 
As one can see, the data are in good agreement with the theoretical prediction. 

\begin{figure}[h]
\begin{center}
  \includegraphics[scale=0.75]{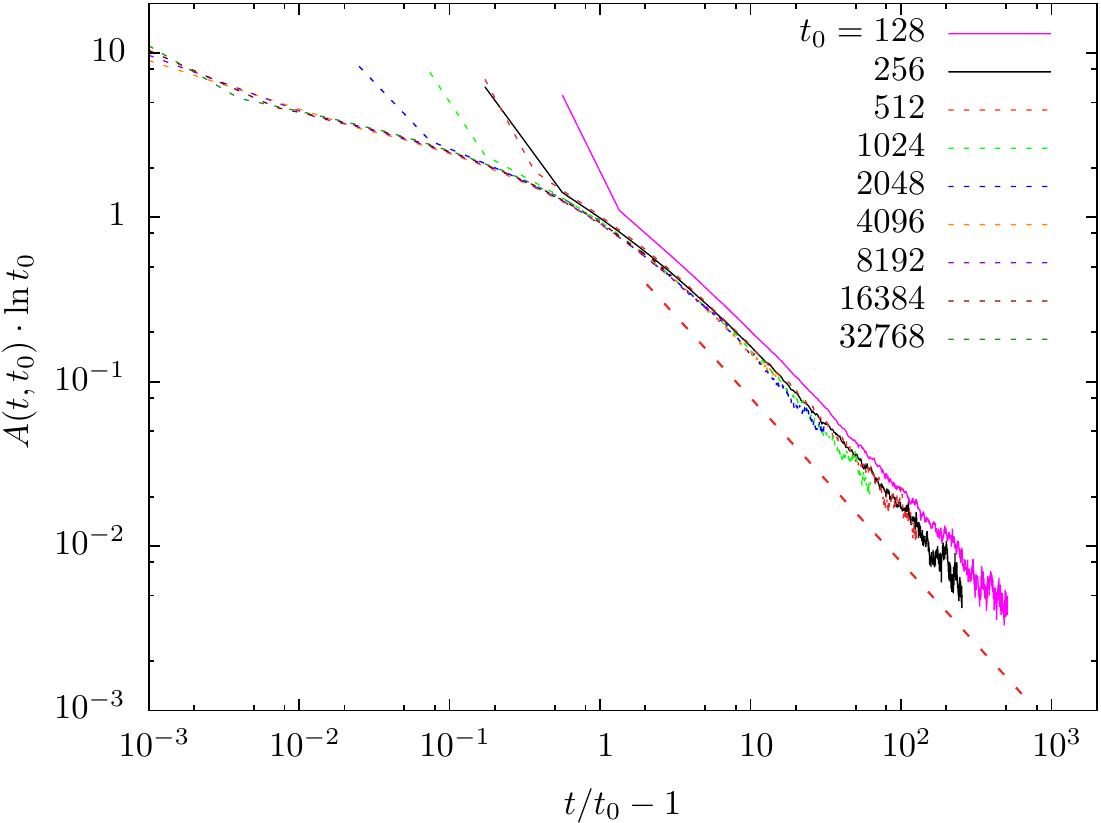}
  \end{center}
  \caption{\footnotesize{ The two-time correlation function $ A(t,t_0)$ times $\ln t_0$ against $ (t / t_0 -1)$, 
  for different values of the waiting time 
  $ t_0 $, in a log-log scale.
  A segment with slope $ - 1 $ is shown as a guide-to-the-eye to stress the good agreement with the 
  exponent found analytically. 
  }}
  \label{fig:scaled_AF_L640}
\end{figure}

In  Fig.~\ref{fig:autocorrelations} we plot, instead, the two-time autocorrelation function	
$ A(t,t_0) $ for values of $ t_0 > 0 $, and two different lattice sizes.
At sufficiently long waiting-time $ t_0 $ the curves tend to flatten, losing their decay. 
This is clearer in the left panel where data for the smaller size, $ L=160 $, are shown. 
In this case, all curves reach a plateau for $ t - t_0 \gtrsim 10^4$, signalling that the steady 
state has been reached. Indeed, we have calculated the average consensus time for a system with size 
$ L=160 $ (see Fig.~\ref{fig:av_CT}), and we found ${\mathcal T}_{160} \simeq 6 \cdot 10^4 $, 
which is compatible with the behaviour of the autocorrelation function.
The same feature is expected to arise for the larger size $ L= 640 $ at a still longer time delay.
In the case $ L = 640 $ we scaled the data to the analytic form (\ref{eq:asympt_final_two_time_corr})
by plotting $\ln t_0 \ A(t,t_0)$ against $t/t_0-1$ in Fig.~\ref{fig:scaled_AF_L640}. The scaling is very good 
for $t_0 \geq 256$.

\vspace{0.25cm}

The numerical analysis of several averaged correlation functions that we have presented so far
is in good agreement with the theoretical predictions for infinite size systems recalled in 
Sec.~\ref{sec:analytic}. However, we will see in the following part of this Section that 
by studying other geometrical observables, we get access to aspects of the dynamics that remain
hidden in the correlation functions. This analysis will allow us to uncover another dynamic regime.

\subsection{Averaged number of wrapping clusters}
\label{subsec:wrapping_domains}

\begin{figure}[h]
\centering
\captionsetup[subfigure]{labelformat=empty}
        \subfloat[]{%
	        \centering
                \includegraphics[scale=0.72]{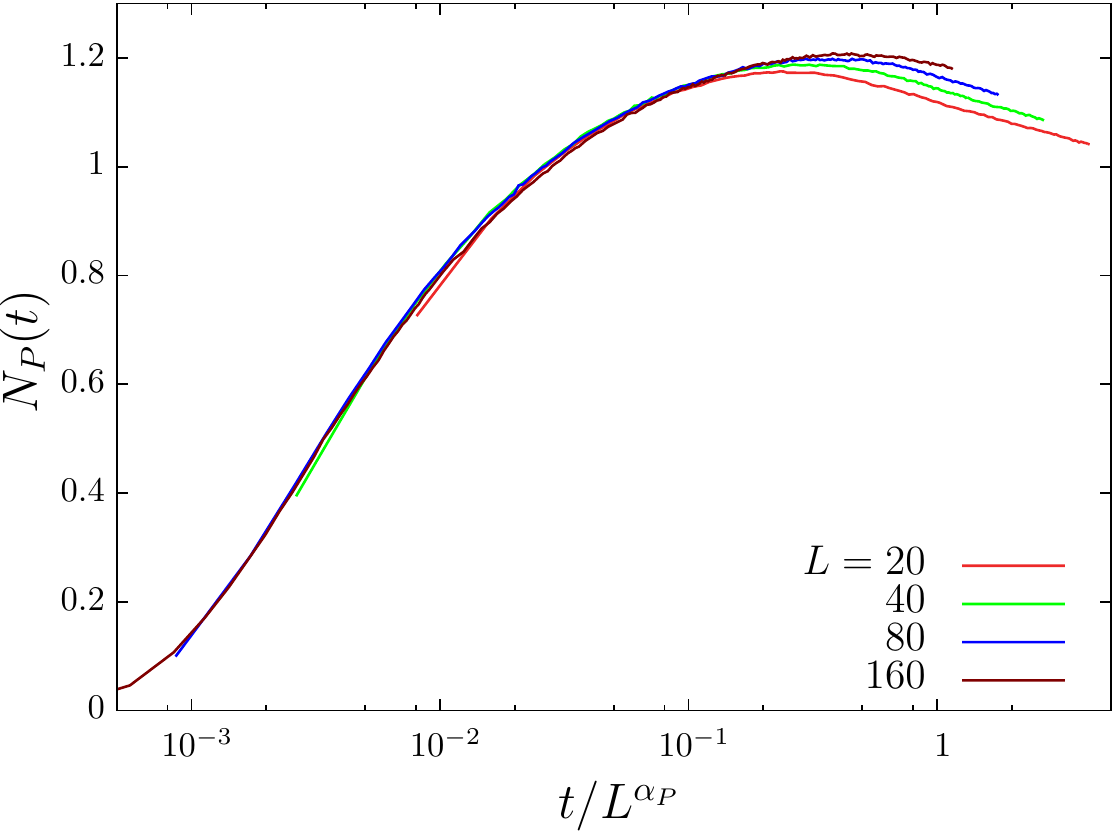}
	}%
        \caption{\footnotesize{ Average number of wrapping domains $N_P(t,L)$ against rescaled time $ t / L^{\alpha_P}$,
        with $\alpha_P=1.667$,  for systems of sizes $ L= 20, \ 40, \ 80, \ 160 $. 
        As $ t \rightarrow + \infty$  the curves should converge to $ 1$ (single domain state).
        }}
\label{fig:wrappings}
\end{figure}

We analysed the time dependence of the average number of wrapping domains per sample, $ N_P(t;L)$,
by supposing that it only depends upon a scaling variable,
\begin{equation}
 N_P (t;L) = \mathcal{N} \left( \frac{t}{L^{\alpha_P}} \right)
 , 
 \label{nwrpa_scaling}
\end{equation}
with $ \alpha_P$ a parameter to be determined and ${\cal N}(u)$ a scaling function. 
At strictly zero argument $ \mathcal{N}(0)=0 $ since the initial random configuration is below the critical 
percolation threshold. At infinite value of the argument $\mathcal{N}(\infty) \to 1$ since the final state of full consensus has a  single domain.
In order to estimate $ \alpha_P$ we tried to collapse $ N_P(t;L)$ for different values of $ L$ by plotting it against
the rescaled time $ t / L^{\alpha_P}$ with trial values of $\alpha_P $. The values of $ \alpha_P$ that gave the best collapse were
found in the range $ 1.65 - 1.68 $, and in Fig.~\ref{fig:wrappings} we present the case
\begin{equation}
\alpha_P=1.667
\; .
\label{eq:alphaP} 
\end{equation}
Even though the quality of the collapse for $ t / L^{\alpha_P} < 0.1 $ is very good, 
we must point out that this is a quite rough estimate, since we could not appreciate remarkable differences between slightly different
values of $ \alpha_P $ in the aforementioned interval.
Deviations from the desired scaling form
$ N_P (t;L) \sim \mathcal{N} \left( t /  L^{\alpha_P} \right) $ are observed for $ t / L^{\alpha_P} \gtrsim 0.1 $. These  are indeed 
expected since the system enters the next dynamic regime of approach to full consensus.
As $ u$ increases from zero, $ \mathcal{N}(u) $ increases monotonically up to a certain value greater than $ 1$. At this stage of the dynamics
there are, then, 
states with more than one wrapping cluster. The scaling function next 
decreases converging to $ 1$ from above. The exponent $ \alpha_P$  sets the typical time
required for the system to reach a regime with wrapping clusters to $ t_P \sim L^{\alpha_P}$.

\vspace{0.5cm}

\subsection{The largest cluster}
\label{subsec:largest-cluster}

We identified the largest cluster at each step of evolution and we computed several of its properties.
This analysis allows us to distinguish whether the largest cluster has 
wrapped around the system in one or more directions and, moreover, to which kind of criticality it belongs.

We first measured the averaged number of its interfaces with positive, vanishing and negative curvature, $\langle N_+ \rangle$, 
$\langle N_0 \rangle$ and $\langle N_-\rangle$, respectively. A non-percolating cluster has a single external interface with 
positive curvature; we therefore call $\langle N_{ep}\rangle = \langle N_+\rangle$ (with $_{ep}$ for external perimeter). 
A cluster that percolates in one direction has two interfaces with vanishing curvature. The interfaces with negative 
curvature are internal to the cluster and surround its holes. 

Another interesting observable  is the area  of the largest cluster  $A_c$ that we normalise as $A_c/L^{D_A}$ with 
$D_A$ a fractal dimension that we need to find. We recall that the fractal dimension
of cluster areas in $2d$ site percolation~\cite{Stauffer94} is
\begin{equation}
D^{\rm cp}_A \simeq 1.896
\; . 
\end{equation}

 Finally, we calculated the averaged total length of the boundary $l_c$ as the sum of the length of external and 
internal interfaces described above. We also normalised this length as $l_c/L^{D_H}$ with $D_H$ a fractal dimension.
For the sake of comparison, we recall 
that for $2d$ site percolation the cluster hull fractal dimension~\cite{SaleurDuplantier87,Coniglio89}
takes the values
\begin{equation}
D^{\rm cp}_H=1.750
\; . 
\end{equation}

\begin{figure}[h]
\centering
	        \subfloat[]{%
	        \centering
                \includegraphics[scale=0.58]{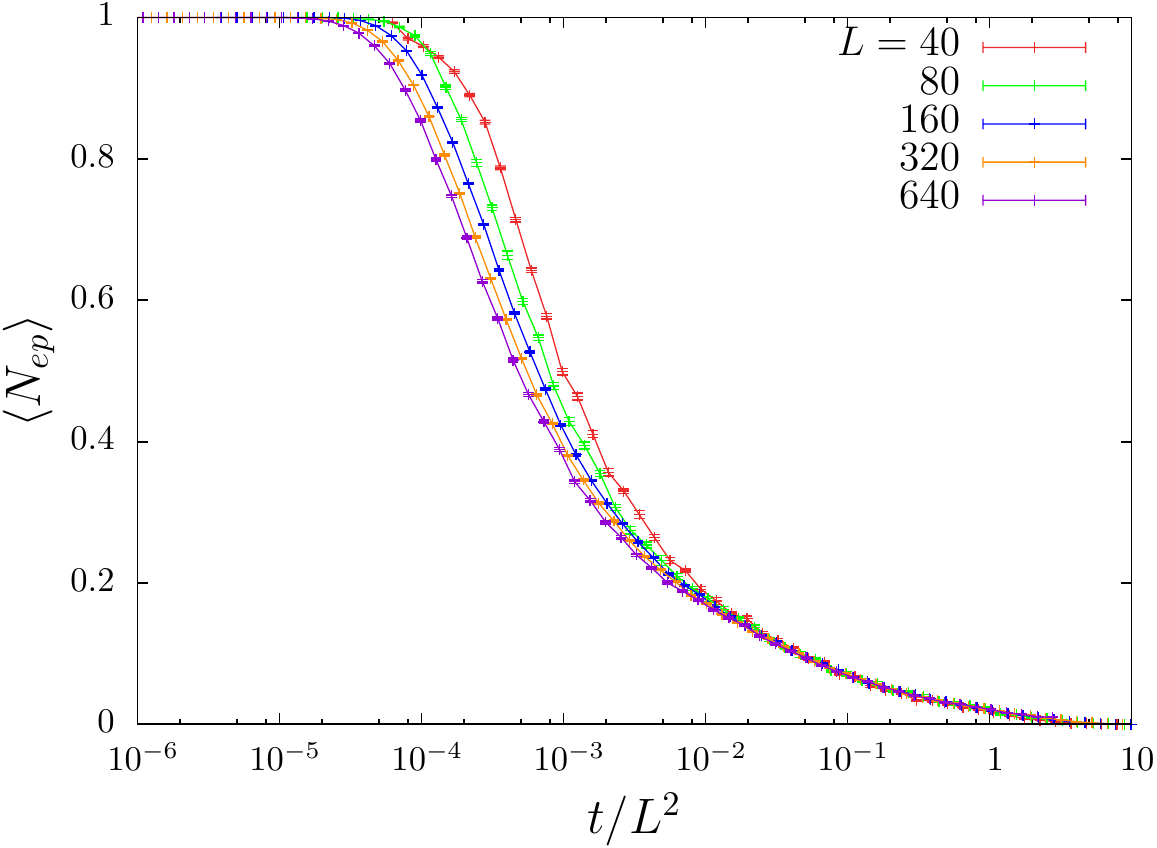}
	        \label{C}
	}%
	\quad
      	\subfloat[]{%
                \centering
                \includegraphics[scale=0.58]{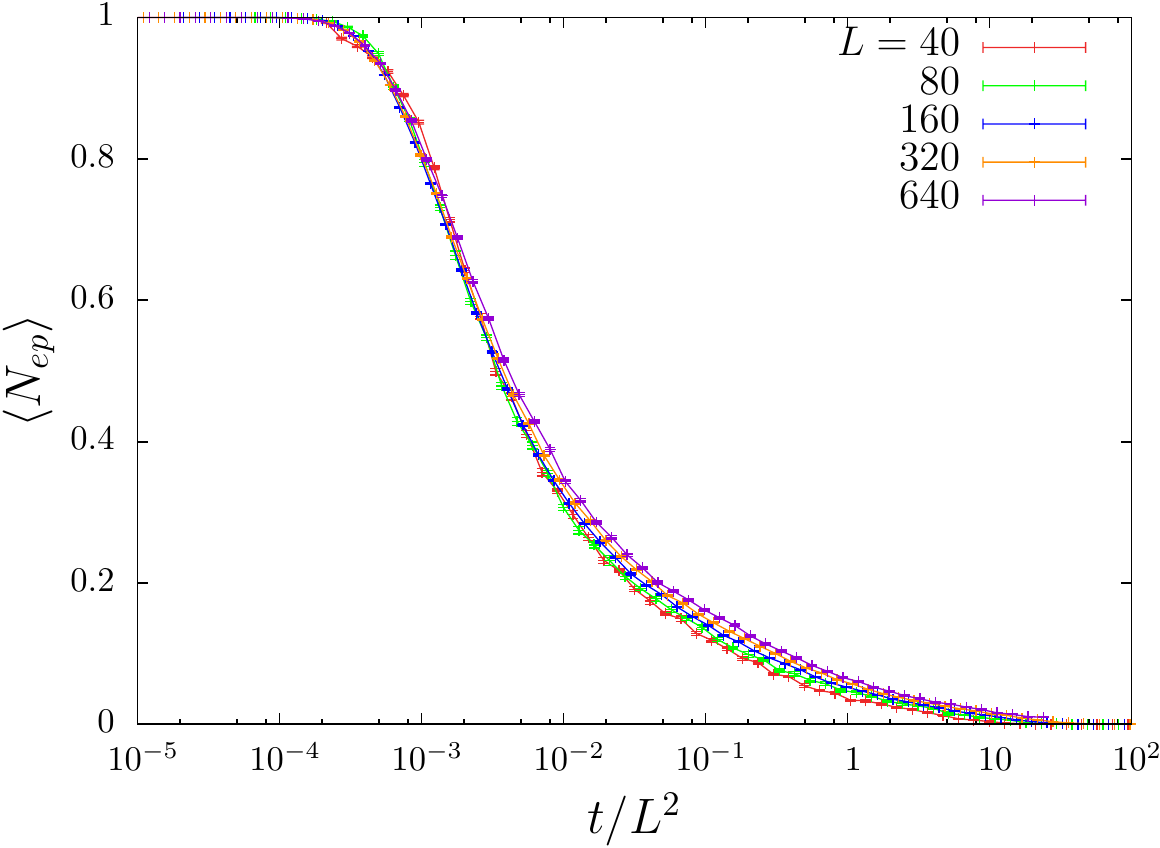}
	        \label{D}
	}%
	\\
	\centering
	        \subfloat[]{%
	        \centering
                \includegraphics[scale=0.58]{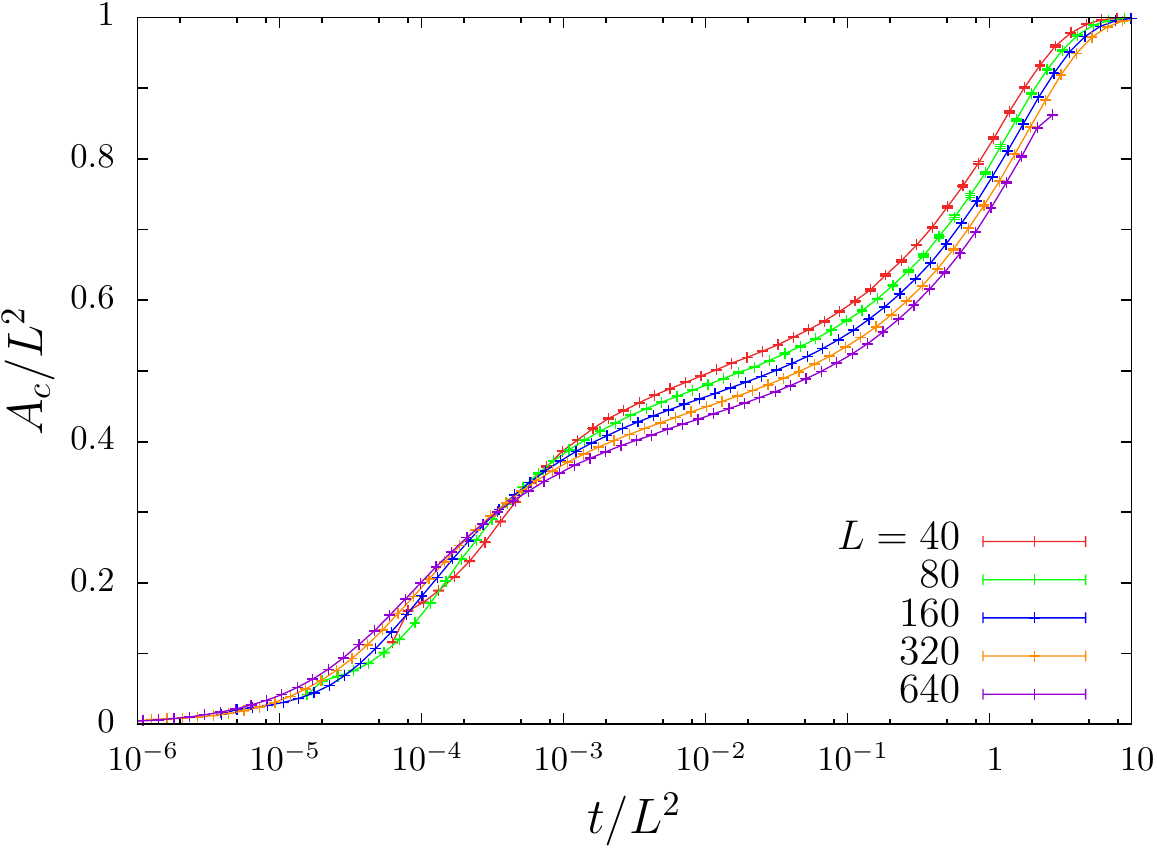}
	        \label{A}
	}%
	\quad
      	\subfloat[]{%
                \centering
                \includegraphics[scale=0.58]{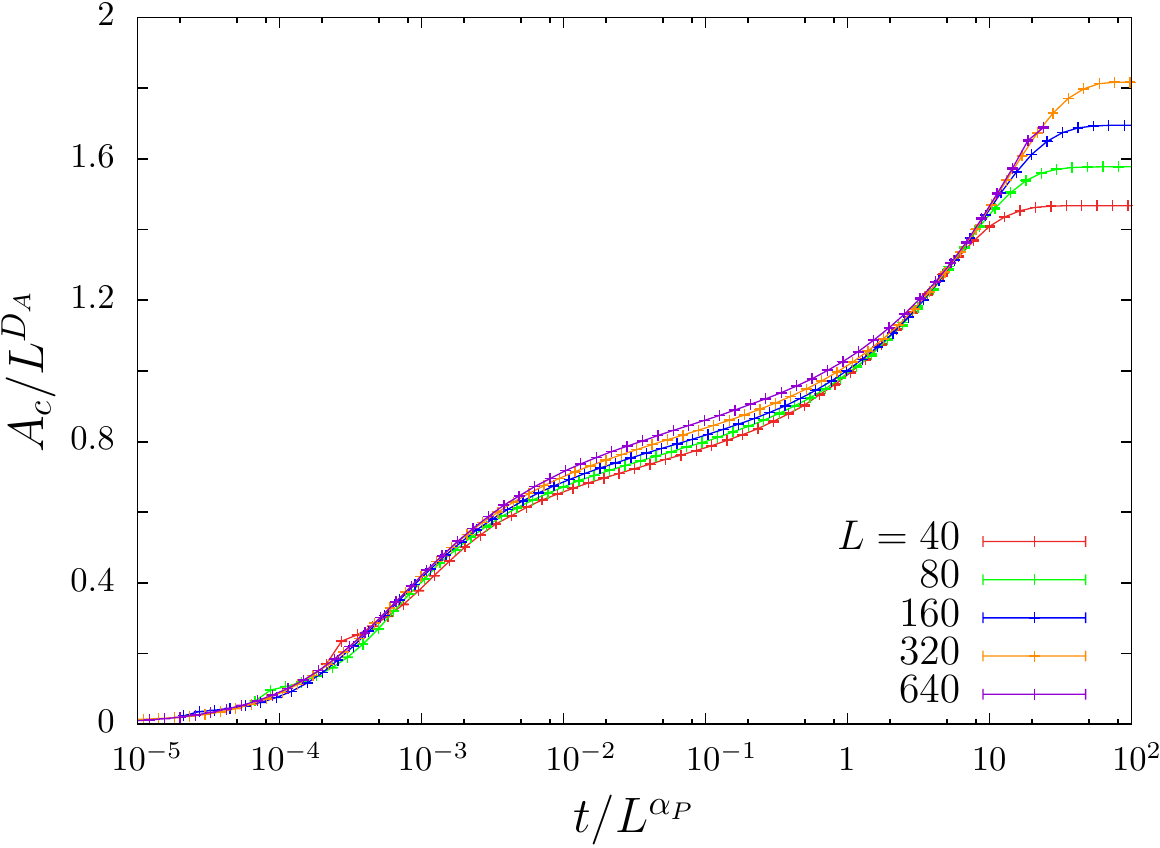}
	        \label{B}
	}%
	\\
	        \subfloat[]{%
	        \centering
                \includegraphics[scale=0.58]{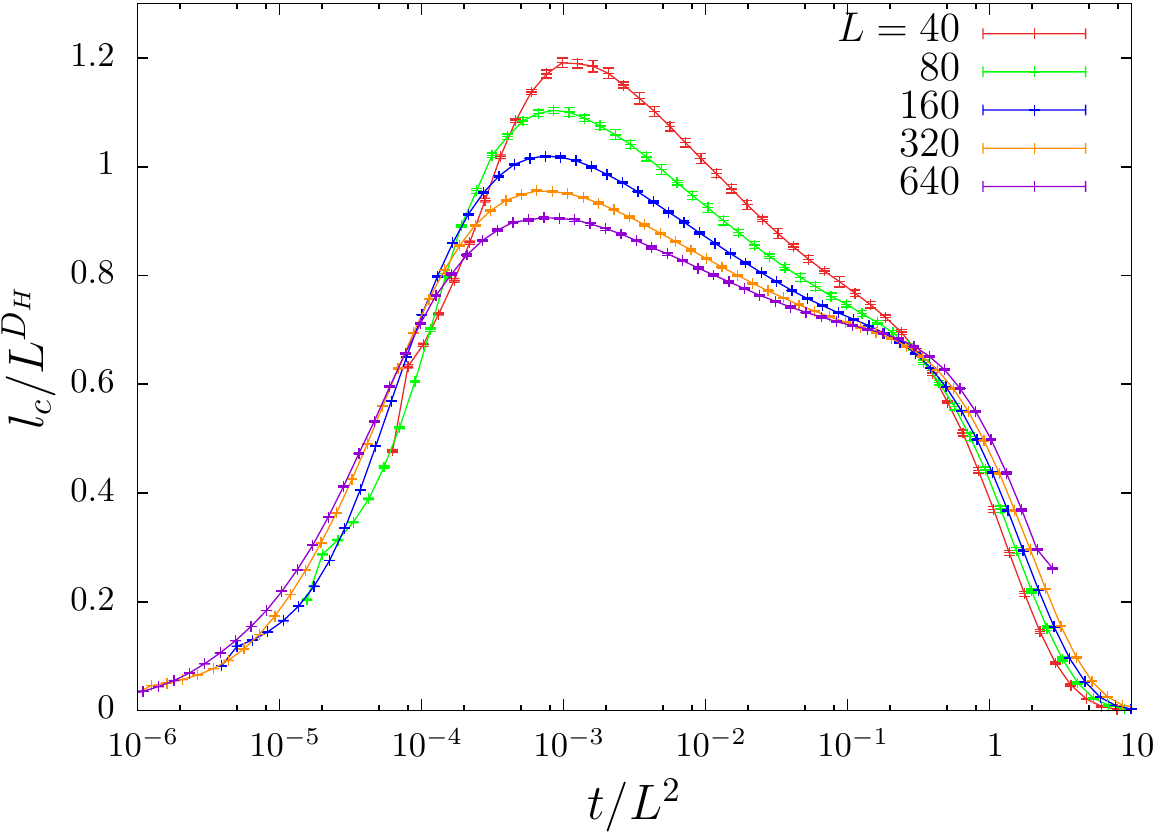}
	        \label{E}
	}%
	\quad
      	\subfloat[]{%
                \centering
                \includegraphics[scale=0.58]{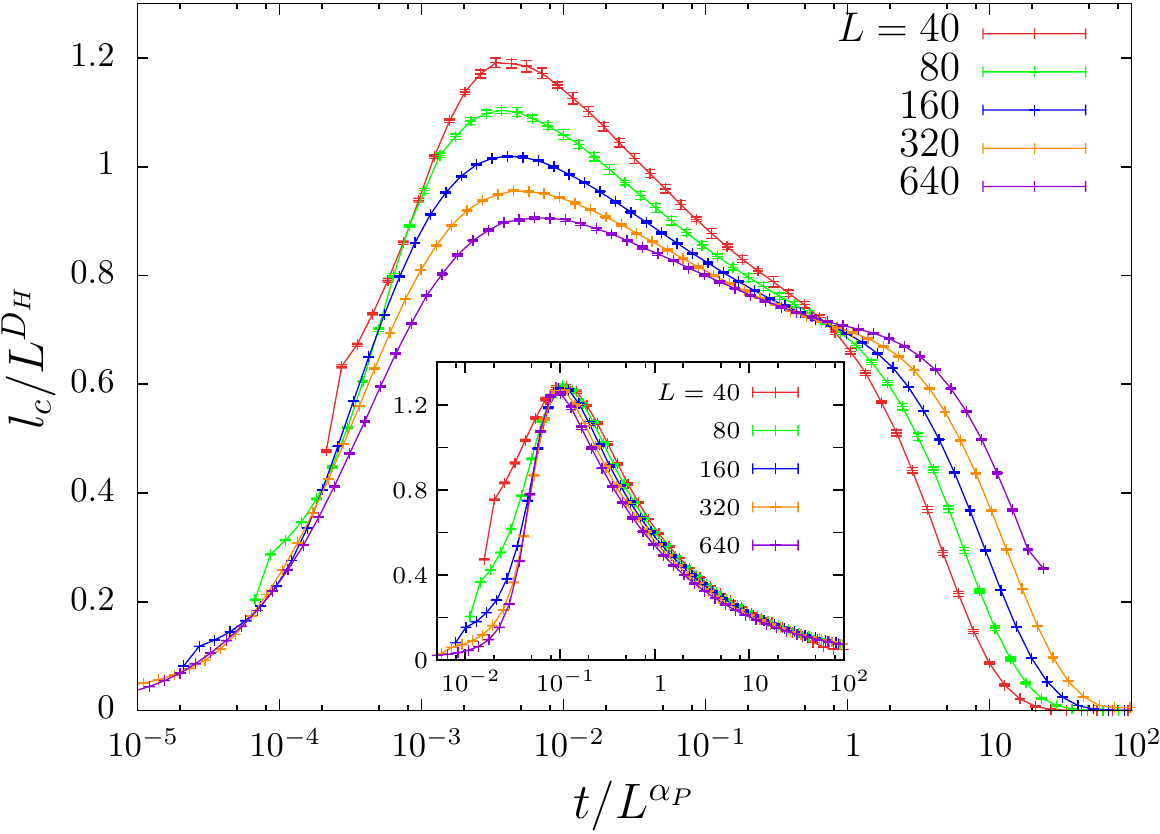}
	        \label{F}
	}%
        \caption{\footnotesize{The largest cluster. The horizontal axes are $t/L^2$ and $t/L^{\alpha_P}$ with $\alpha_P=1.667$
        in all panels in the left and right columns, respectively. 
        (a)-(b) The averaged number of external interfaces (only for non-percolating clusters).
        Its area $A_c$ normalised by $L^2$ in (c) and $L^{D_A}$ with $D_A=1.89583$ in (d).
         (e)-(f) Its total perimeter length in the voter  and $2d$IM quenched  to $T=0$ (inset)
         normalised  by $L^{D_H}$ with $D_H=1.75$.
}}
\label{fig:percolation}
\end{figure}

\indent
All these quantities show different scaling properties at short times before percolation is reached and at longer times, when the percolating 
cluster has established, and the further evolution takes the system to the final absorbing state. In the former time-regime, the data scale as a 
function of $t/L^{\alpha_P}$ with 
$
\alpha_P=1.667
$, see all panels in the right column in Fig.~\ref{fig:percolation}, while in the latter they do as a function 
of $t/L^2$ (apart from logarithmic corrections), see all panels in the left column in the same figure. The value of $\alpha_P$ is rather large in the voter model 
(much larger than in the $2d$IM where $\alpha_P=1/2$~\cite{BlCoCuPi14}) and this makes the distinction between the regime of approach to percolation and the further 
evolution towards full consensus very hard.

The averaged number of interfaces with positive curvature smoothly decays from  one to zero as more and more samples wrap around the sample in 
at least one direction, see Fig.~\ref{fig:percolation}~(a)-(b). Concomitantly, the averaged number of maximal size clusters that wrap around the sample 
in one direction increases in time from zero initially to a value that is close to 0.9 to later decay again to zero when the cluster percolates in the other 
direction as well, leaving only internal interfaces in the system (not shown). The averaged number of internal boundaries has a very similar qualitative 
behaviour to the one of the zero-curvature ones (not shown either). Panel (a) confirms the scaling with $t/L^2$ at long times while panel (b) 
indicates that the good scaling variable at short times is $t/L^{\alpha_P}$.

Figures~\ref{fig:percolation}~(c)-(d) and (e)-(f) display the area  of the maximum size cluster and the total perimeter length, 
respectively. The asymptotic approach of $A_c/L^2$ in (c) to one and $l_c$ in (e) to zero confirm that the systems approach full consensus.
They also prove that there are no blocked states in the voter model as there are in the $2d$IM at $T=0$~\cite{SpKrRe01,SpKrRe02,BaKrRe09,OlKrRe12,BlPi13}.

The scaling of the horizontal axis in panels (c) and (e) is intended to work at long times only. 
The persistent deviations from scaling should be due to the fact that we did not include in the 
 scaling variable corrections that depend logarithmically on the system size, 
although we know from the analysis of the consensus time that these should exist.
 
The scaling of the vertical axis in panel (d) is intended to determine the fractal dimension of the area of the largest cluster,
in the short time regime. We used different values of $D_A$ and we found that the most satisfactory 
collapse in the short-time regime is found for $D_A$ in the interval  $1.89 - 1.93$. In panel (d) we show the scaling
found using the fractal dimension of $2d$ critical site 
percolation, $D_A = D_A^{\rm cp} \simeq 1.89583$ which is within this interval (while $D_A$ for other critical states lies outside this interval, 
{\it i.e.} $D_A^{\rm ci} =1.948$ at the critical Ising point). Once again, due to the large value of 
$\alpha_P$ the two relevant asymptotic regimes are not well separated, and we cannot do better 
than this in the determination of $D_A$. 

Similarly, the scaling of the vertical axis in panel (f) should determine the fractal dimension of the 
boundary of the largest cluster, $D_H$ (although we stress that we show data for the total 
interface length here). Using $D_H=D_H^{\rm cp}= 1.75$ we see that all curves cross at $t/L^{\alpha_P}
\simeq 1$ suggesting that critical percolation is reached at around $t_P \simeq L^{\alpha_P}$. Other choices
for the value of $D_H$ do not allow for data collapse at any value of the scaling variable. Furthermore, 
one could argue that the scaled data for $t/L^{\alpha_P}$ are slowly approaching, for increasing
system size, a flat form.
 
In the inset to panel (f) we show, for comparison, the same scaling plot, $l_c/L^{D_H}$ against $t/L^{\alpha_P}$ for the zero temperature Ising model
quenched from $T_0\to\infty$, 
with $D_H=1.75$ and $\alpha_P=0.5$. Apart from the rather small system sizes, $L=40, \ 80$, the data for the larger 
systems show a very good collapse over a rather large time-window. In the Ising model the two asymptotic 
time regimes are rather well-separated (as $\alpha_P=0.5$ is quite different from $z_d=2$) and this fact contributes
to the good data collapse. 

\subsection{Two growing lengths}
\label{subsec:two-growing-lengths}

\par
We conclude that, as in the $2d$ Ising model with non-conserved order parameter~\cite{ArBrCuSi07,SiArBrCu07,Blanchard,BlCoCuPi14},
the dynamics of the finite size $2d$ voter model takes place in two distinct regimes: the system first develops wrapping clusters 
of  critical  site percolation kind; once these
are established, the further growth is a more usual coarsening process. The two growing lengths controlling the 
evolution in the two regimes are 
\begin{equation}
\ell_P(t) \simeq t^{1/\alpha_P}
\; , 
\qquad\qquad
\ell_G(t) \simeq t^{1/z_d}
\; . 
\end{equation}
In the Ising model on a square lattice the exponents $\alpha_P$ and $z_d$ are rather different, $\alpha_P=1/2$~\cite{BlCoCuPi14}  and $z_d=2$. Since $\alpha_P$ is so small, 
for the system sizes used in numerical simulations the approach to percolation occurs in a few steps
and the time window is not sufficiently long to allow for a careful dynamic scaling analysis. Instead, in the voter model,
$\alpha_P\simeq 1.667$ is quite large and not very different from $z_d=2$. The system takes much longer to reach critical percolation
and the advantage is that a rather wide time window can be explored in which the relevant growing length is $\ell_P(t)$. 

\subsection{Density of domain areas}
\label{subsec:density-domain-areas}

We now turn to the statistics of domain areas. We recall that we defined a cluster or a domain as the connected ensemble of 
nearest-neighbour parallel spins and its area as the number of spins in it. 

\begin{figure}[b]
\centering
	{%
                \centering
                \includegraphics[scale=1.2]{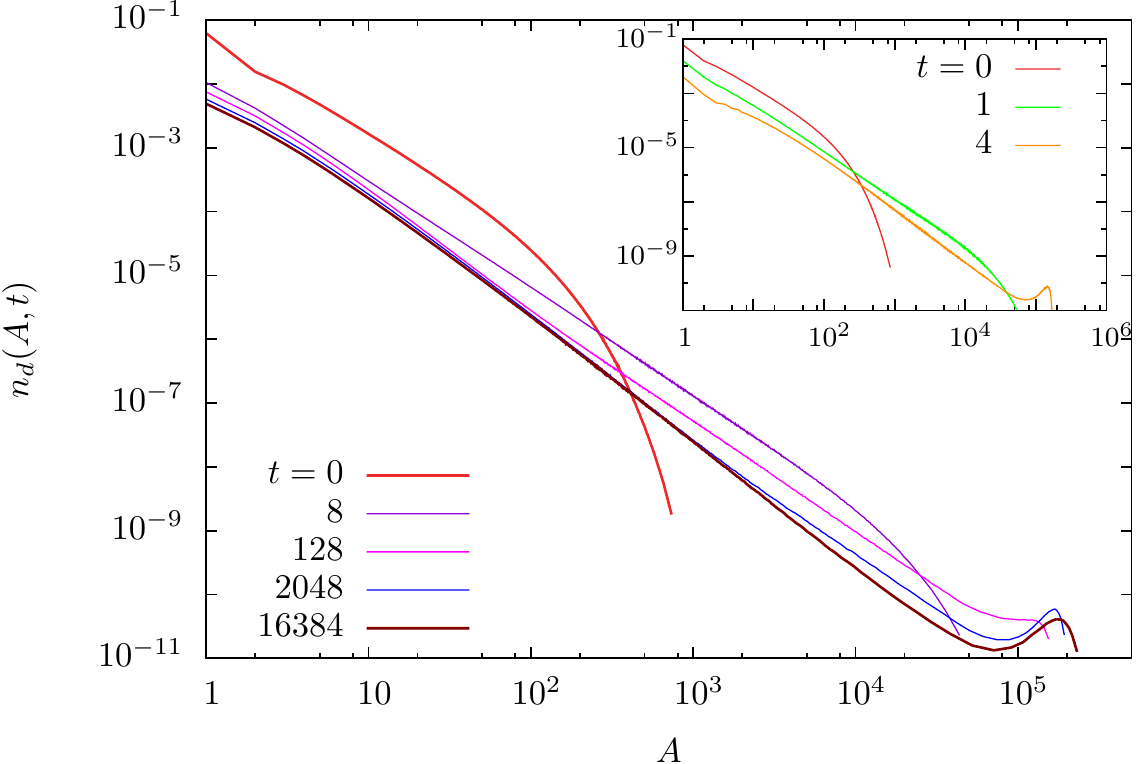}
	}%
        \caption{\footnotesize{Instantaneous domain area number densities.
        Main plot:  $2d$ voter model with linear size $L=640$. 
	Inset: $2d$IM quenched to $T=0$ from $T_0 \to\infty$ and $L=640$.
	In both cases the curves are presented in a double logarithmic scale
	and the times at which the data are collected are given in the key. }}
\label{fig:domain_numb_density}
\end{figure}

Figures~\ref{fig:domain_numb_density} shows raw data for the time-dependent number density of domain areas in 
systems with linear size  $L=640$. 
Initially, the curves show no particular structure, as the random initial condition is non-critical and the weight of the 
distribution at large areas drops significantly. 
However, as time elapses, a power law extending over several decades
develops, as already noticed by Scheucher and Spohn~\cite{ScheucherSpohn88}. Moreover, a bump with support over areas that are of the order
of magnitude of the size of the system also appears. These areas are the ones of clusters that wrap around the 
system, as the ones discussed in the previous subsection. The height of the bump increases in time. It 
tends to become stable at the longest time-scales used, $t \simeq 17 \cdot 10^3$. (For a smaller system size, say $L=160$, the same
features are realised but, in contrast, after growing in height the bump tends to wash out, 
after times of the order of $ 2 \cdot 10^3$.)
 This feature is very similar to what was observed in the $2d$IM quenched to zero temperature, although a stable bump, linked to 
 the system reaching critical percolation, establishes at a much shorter 
  time-scale, $t_P \simeq 5$ for similar system sizes~\cite{ArBrCuSi07,SiArBrCu07} as $\alpha_P =1/2$ in this case~\cite{BlCoCuPi14}. Indeed, 
  the curves for $t=0, \ 1, \ 4$ in the inset are qualitatively identical to the ones for $t=0, \ 8, 16384$ in the main plot. (We will make a quantitative 
  comparison between the behaviour in the two models below.)
In the case of the $2d $ voter model we observed that percolating clusters can appear in early stages of the dynamics, but
they tend to break soon after their formation and reappear later on, taking longer to establish a stable pattern, see Fig.~\ref{fig:pc_snapshots}. In particular, for a system with linear size
$L=640$, a stable structure of percolating domains  establishes only after a time of the order of $ t_P \sim 10^5$.

\begin{figure}
\centering
                \centering
                \includegraphics[scale=1.1]{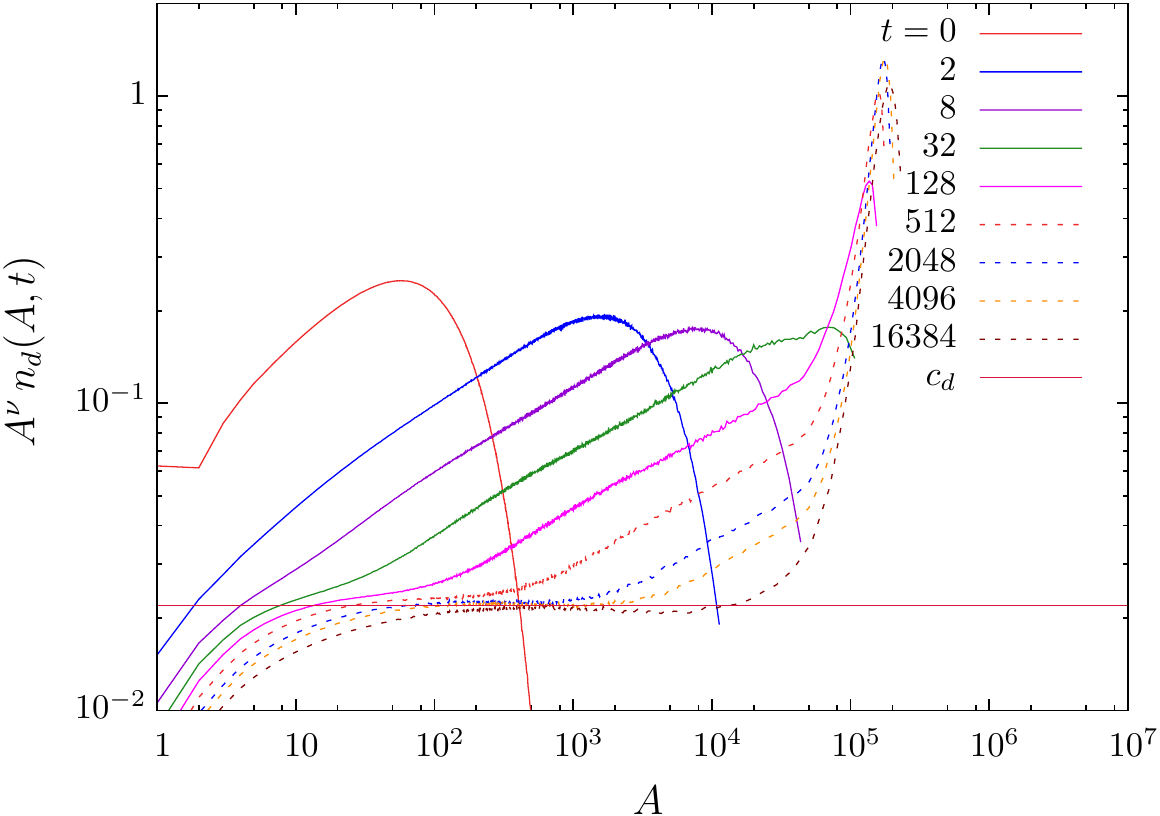}
	        \label{B}
\caption{\footnotesize{ Time-dependent number density of domain areas, $n_d$, multiplied by $A^\nu$ with 
	$\nu \simeq 1.98$ for $L=640$. As the system comes closer to the percolating state, the number density of domain 
      	areas tends to the time-independent  form $ n_d(A,t) \sim c_d \ A^{\nu}$, save for a correction due to wrapping domains in the bump.
	The value of $\nu$ was estimated by fitting the aforementioned functional form to the data corresponding to the latest time. 
	The horizontal line is at $c_d \approx 0.022 $.
}}
\label{fig:domain_power-law}
\end{figure}

The analytic and numeric analysis of the $2d$IM quenched from infinite to zero~\cite{ArBrCuSi07,SiArBrCu07} or the critical~\cite{Blanchard} temperature showed
that the number density of areas approaches a scaling form 
\begin{equation}
A^{\nu} \ n_d(A,t) = \Phi \left(\frac{A}{\ell^{D_A}(t)}\right)
\label{eq:critical-scaling}
\end{equation} 
where $D_A$ is the fractal dimension of the areas studied, $\ell(t)$ is the relevant growing length
and $\Phi$ a proper scaling function. 
After $t_P$ this scaling has to be corrected
by an additive term that takes into account the percolating clusters that had already established (the bump).
In the zero temperature quenches of the $2d$IM the approach to percolation was so fast that the study of this scaling for times such that 
the relevant growing length is $\ell_P(t)$ was not performed. In the critical quenches in~\cite{Blanchard} a triangular lattice for which the system was already at the critical 
percolation point initially was used. Here, with the voter model, we have the possibility of studying  the dynamic scaling in the regime of 
slow approach to percolation in detail, by taking advantage of the large value of $\alpha_P$.

The same datasets used in Fig.~\ref{fig:domain_numb_density} are plotted in the form $A^\nu \, n_d(A,t)$ against $A$ in 
Fig.~\ref{fig:domain_power-law} with 
$\nu \simeq 1.98$ for $L=640$. 
The value of the exponent $ \nu$ was found by fitting the data $ n_d(A,t)$ corresponding to the longest time reached in the simulation
($ t=2048$ for $ L=160$, not shown, and $ t=16384$ for $L=640$) with the power law $c_d A^{-\nu}$, in the range of areas
[$10^{2}$, $10^{3}$] for $L=160$ and [$10^{2}$, $10^{4}$] for $L=640$.
We found  $c_d = 0.0207 \pm 0.0001 $ and $\nu= 1.972 \pm 0.005 $ in the case $L=160$, and 
 $c_d = 0.0220 \pm 0.0001 $ and $\nu= 1.980 \pm 0.001 $ for $L=640$.

The value of the exponent $\nu $ increases very weakly with $L$ and should be larger than $2$ in the infinite
size limit to ensure that the average area of non-percolating domains, $\int dA \ A \  n_d(A,t) $, is a finite quantity.
However, the approach to the asymptotic limit is so slow that it is very hard to get closer to it numerically.
This particular feature of the dynamics was also observed in~\cite{ScheucherSpohn88} for the voter model and 
we stress that, in the $2d$IM, the expected value $\nu=2.05$ is found only for a very careful choice of the 
areas to fit.
 
\begin{figure}[h]
\centering
      	\subfloat[]{%
                \centering
                \includegraphics[scale=1.2]{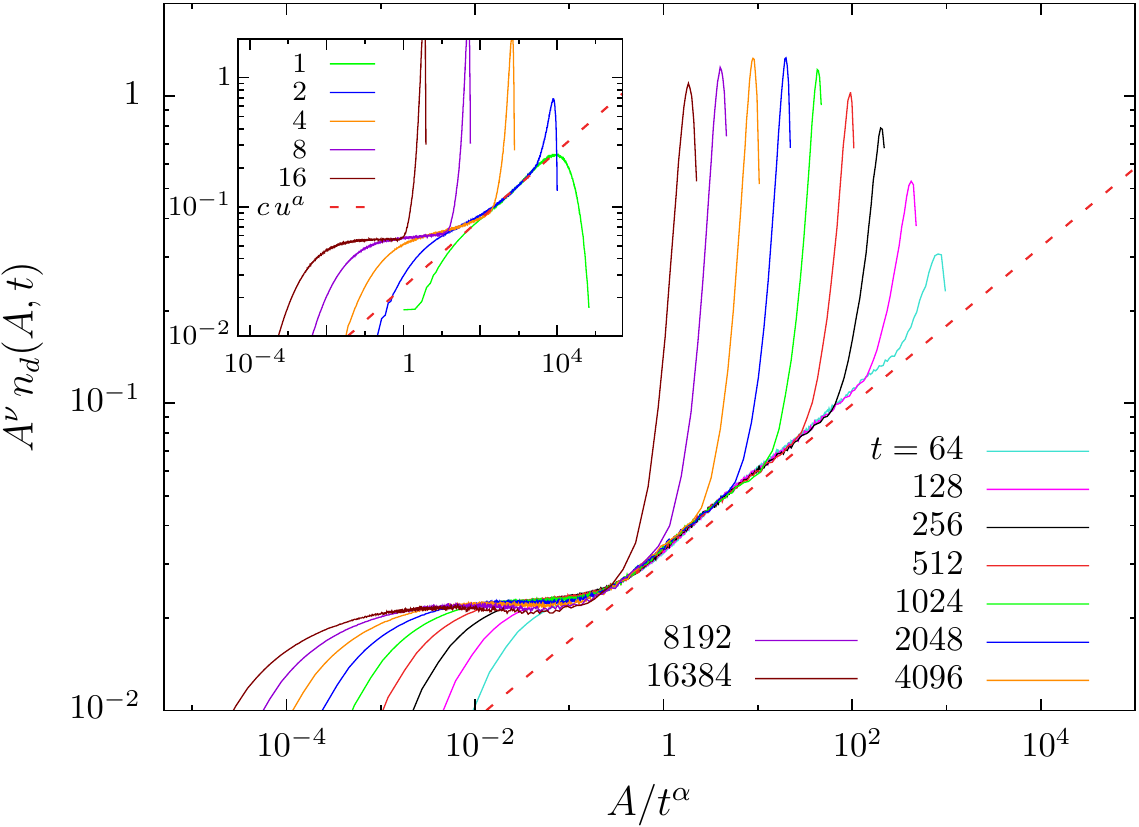}
	        \label{B}
	}%
        \caption{\footnotesize{Number density of domain areas multiplied by  $A^\nu$ against the rescaled area $ A / t^{\alpha} $, for
	$ L=640 $. The  exponent $ \nu$ takes the same values as in Fig.~\ref{fig:domain_power-law}:
        $ \nu \simeq 1.98$. The value of $ \alpha $ that yields the best time-scaling is $\alpha \simeq 1.19$.
        A fit to the function $ f(u) = c \ u^a$  of the data at $t=64$ over the interval $[1, 100]$ yields
        $ a \approx 0.26$. The red dashed line was added to better visualise this power-law behaviour. Inset: 
        the same scaling plot for the $2d$IM after a quench to $T=0$, at five short times given in the key. 
        The plateau is at $2c_d \simeq 0.044$ as explained in~\cite{ArBrCuSi07}.
        The power law shown with a dashed line was drawn with the same value of the exponent $a$ as for the 
        voter model.  
}}
\label{fig:domain_timescale}
\end{figure}

The value taken by the constant $c_d$ is very relevant to our discussion. Indeed, it was used in~\cite{ArBrCuSi07,SiArBrCu07} to 
distinguish the criticality of large scale domains in zero-temperature 
quenches of the $2d$IM from equilibrium at $T_0\to\infty$ and $T_0=T_c$. More precisely, 
the area distribution of clusters of occupied sites at critical  site percolation and, say, domains of positive spins at the critical Ising 
point are given by $n_d(A) \simeq 2C/A^{\tau^{\rm cp}}$ and $n_d(A,0) \simeq C/A^{\tau^{ci}}$, respectively, with 
$2C \simeq 0.023$, and $\tau^{\rm cp} =  1 +d/ D^{\rm cp}_A$ and $\tau^{\rm ci} =  1 +d/D^{\rm ci}_A$ the Fisher exponents 
related to the fractal dimensions of the domain areas 
under the two critical conditions. Cardy and Ziff~\cite{CaZi03} obtained these universal constants
analytically for the number density of {\it hull-enclosed} areas instead of {\it domain} areas using a Coulomb gas approach.
Arguments presented in~\cite{SiArBrCu07} suggest that very close values should apply to domain areas as well. 
Numerical simulations on the square and triangular lattice confirmed 
the value obtained with field theoretic methods for hull-enclosed and domain areas~\cite{CaZi03,SiArBrCu07,BlCoCuPi14} 
After a zero-temperature quench of the $2d$IM with
initial states drawn from infinite temperature and critical temperature conditions, the evolving large scale areas are distributed 
algebraically and the number densities have Fisher exponents {\it and}  constants in  the numerator that are the ones 
cited above for critical  site percolation (see the inset to Fig.~\ref{fig:domain_timescale}) 
and critical Ising conditions, though both multiplied by a factor of two
when clusters of both (up and down) species are counted~\cite{SiArBrCu07}.

In the voter model with random initial conditions we find $c_d \simeq 0.022$ that 
is consistent with $c_d=2C$  (within numerical accuracy), see Fig.~\ref{fig:domain_numb_density}, 
and with critical Ising initial configurations we find a constant taking the value $c_d/2=  C$ (not shown). This result confirms 
the reduction of the number density of finite areas by two for initial conditions with long-range correlations
with respect to the ones with only short-range correlations.

In Fig.~\ref{fig:domain_timescale} we show  $ A^{\nu} \, n_d(A,t)$ 
against the rescaled area $ A / t^{\alpha} $ for systems with $ L=640$. 
The value of $ \alpha $ that allowed us to obtain the best collapse was found to be approximately 
equal to $ 1.19 $.  This value is in good agreement with the prediction
$\ell_P^{D_A}(t) \simeq t^{D_A/\alpha_P} = t^{1.90/1.67} = t^{1.14}$. In the inset we perform the same 
analysis on the $2d$IM quenched to $T=0$, by focusing on the very short time dynamics such that $t \ll t_P$.
In agreement with the proposal, the curves collapse if one uses $A/t^{D_A/\alpha_P} = A/t^{(187/96)/0.5} = A/t^{3.9}$.

Apart from deviations caused by the appearance of wrapping domains in the bump, for large values of $ t $
all curves seem to have the same behaviour, namely two distinct regimes: for $ A / t^{\alpha} \lesssim 0.1$ there is a nearly flat region, which would
mean that $ \Phi(u) \sim$ const. and thus $ n_d(A,t) \sim A^{-\nu}$, {\it i.e.} the statistics of domain areas is independent of time;
for $ A / t^{\alpha} \gtrsim 0.1 $ instead, the scaling function seems to behave as an increasing power law 
$ \Phi(u) \sim c \, u^a $, with a self-similar statistics of domain areas 
in the sense that it depends only on $ A / t^{\alpha}$.
A fit of the data for the shortest time in Fig.~\ref{fig:domain_timescale}
on the interval $ [ 1, 100]$ of the scaling variable $ u=A /t^{\alpha}$, yields 
$ c = 0.034 \pm 0.001 $ and $ a = 0.257 \pm 0.001 $. Analogously, for the case $ L=160$ 
we obtained  $ c = 0.032 \pm 0.001 $ and $ a = 0.264 \pm 0.001 $ (not shown).
The existence of two distinct regimes for small and large values of $ A / t^{\alpha}$ is also observed in the Ising 
model as shown in the inset of Fig.~\ref{fig:domain_timescale}. Moreover, for large values of $ A / t^{\alpha}$
we also observe an increasing power law (also shown in the inset) with a very similar power, 
{\it  {\it i.e.}} $ c = 0.024 \pm 0.001$ and $a=0.262 \pm 0.001 $ for $L=640$.

\par
Coming back to the domain area statistics, as one can see from the plots, the flat region for $ A / t^{\alpha} \lesssim 0.1 $ 
becomes larger as time increases, while the complementary region of larger domains  shrinks, until disappearing.
Note that as time increases
the percolating (wrapping) domains become more and more predominant and eventually the number domain density converges
to the absorbing state form which is just a delta function centered at $ A = L^2$.
Even though this fact does not rule out the possibility of a transient regime in which more than one stable percolating clusters coexist, similarly
to what happens in the zero-temperature 2$d$IM on a finite lattice, we found
that it establishes during a very short time period (compared to the whole duration of the dynamics) before the consensus state is reached, so
it is somehow difficult to ``catch'' it in the domain area statistics.

\subsection{Space-time correlation function}
\label{correlation_function}

\begin{figure}
\centering
        \subfloat[]{%
	        \centering
                \includegraphics[scale=0.65]{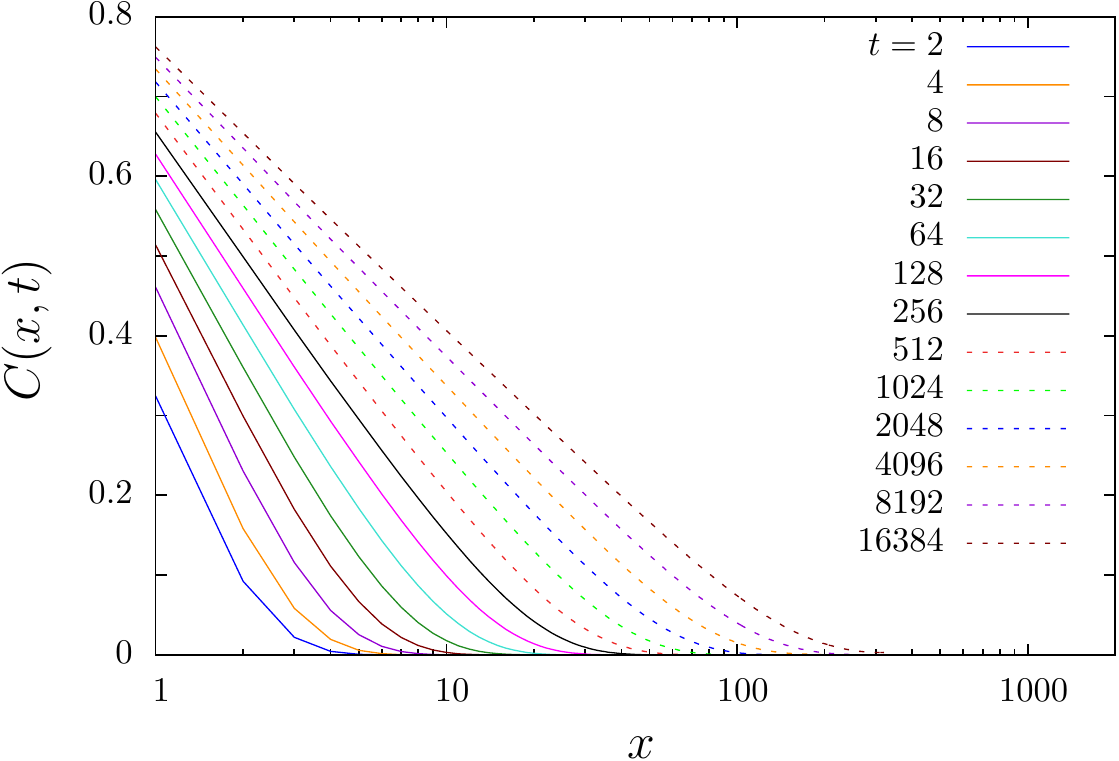}
	        \label{A}
	}%
	\quad
      	\subfloat[]{%
                \centering
                \includegraphics[scale=0.59]{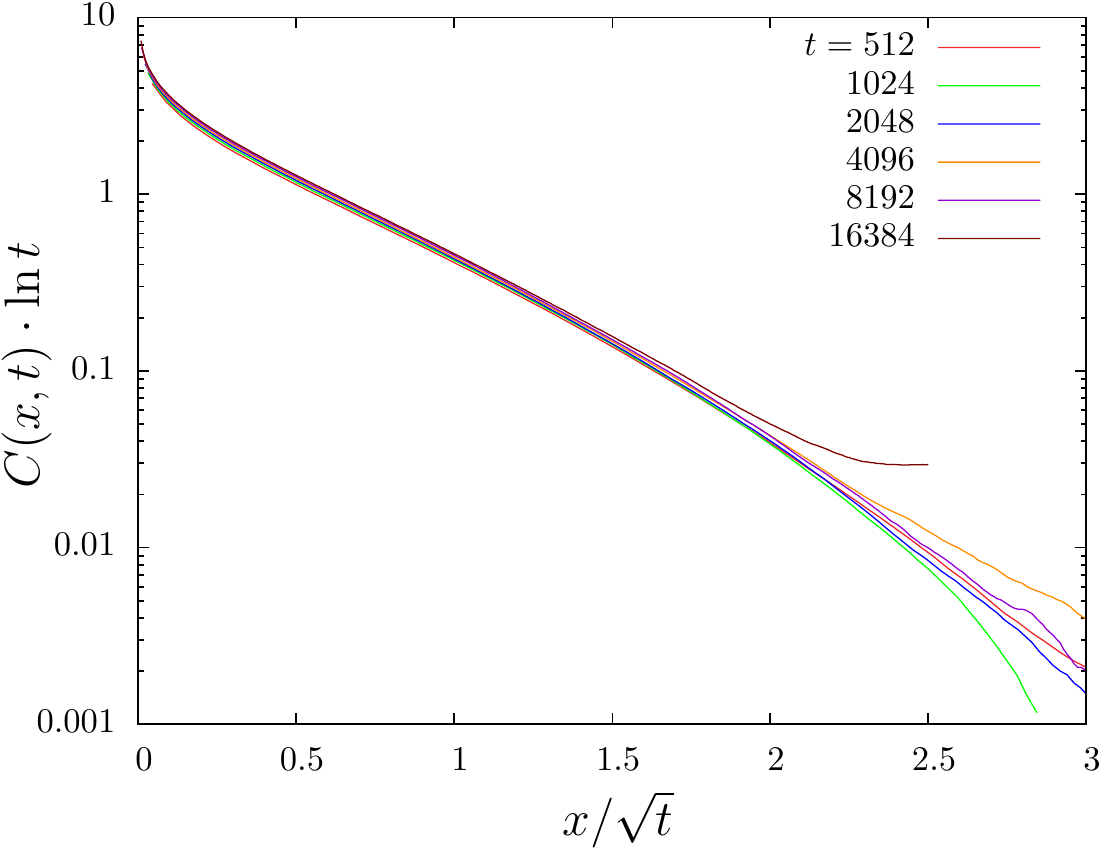}
	        \label{B}
	}%
        \caption{\footnotesize{In panel (a) log-linear plot of the space-time correlation function along a principal direction of the lattice, {\it i.e.}
 $ C(x,t) = {\langle s_{x \boldsymbol{\mathrm{e}}_i}(t) s_{\boldsymbol{0}}(t) \rangle} $,
        at different times for a system with linear size $L=640$. 
In panel (b) the same data as in panel (a)  multiplied by
$ \ln{ t }$ and plotted against the scaled distance $ x/\sqrt{ t} $ in linear-log scale. See the main text for a discussion.
        }}
\label{fig:correlf_L640}
\end{figure}

Having established the existence of two dynamic growing lengths in a finite size system, we now
put the scaling form of the space-time correlation, Eq.~(\ref{correlf_asympt3}), to the numerical test.
Figure~\ref{fig:correlf_L640}~(a) shows data for $C(r,t)$ on a lattice of linear size $L=640$.
The correlation function was calculated only along a principal direction
of the lattice (e.g. the horizontal direction), as 
$C(\boldsymbol{x},t)$ should be isotropic and depend on $\boldsymbol{x}$ only
at  distances much longer than the lattice spacing.
In Fig.~\ref{fig:correlf_L640}~(b) the correlation function multiplied by $ \ln{  t} $ is plotted against the scaled 
distance $ x/\sqrt{ t} $. The curves at different times tend to collapse even though they deviate for 
large values of $ x/\sqrt{ t}$. These deviations are due to finite size effects: 
since we have taken periodic conditions at the boundaries, the
data at distances  of the order of the lattice size, specifically $ x \simeq L / 2$, 
are much affected by the boundaries. One reckons that, consistently, the deviation from the scaling law for
large $ x/\sqrt{ t}$ occurs at smaller values of the scaling variable at longer times.

\par
However, from the analysis of the clusters we now known that in the dynamic regime in 
which percolating clusters develop there is another characteristic length in the problem,
$\ell_G(t)$. Accordingly,
the scaling form of the correlation function has to be modified to capture the dynamics in both dynamic regimes
(see~\cite{BlCoCuPi14} for this analysis in the $2d$IM).
We therefore introduce a new two-variable scaling function $ g(u,v) $
\begin{equation}
C(\boldsymbol{x},t; L) = \frac{1}{\ln{ (  t/\tau ) }} \; 
g \left( \frac{ | \boldsymbol{x} |}{\ell_G(  t/\tau )} \, , \; \frac{\ell_P(t/\tau)}{L} \right)
\; , 
\label{CF_new_scaling_form}
\end{equation}
such that for $t\simeq t_P \simeq L^{\alpha_P}$, the new scaling variable is close to one, $\ell_P(t)/L \simeq 1$, 
and one recovers the infinite size limit.
This suggests that the data for $ C(x,t;L) $ at different times $ t_i$ 
and sizes $ L_i$ chosen in a such a way that the ratio $ \chi = \ell_P(t_i)/L_i $ is kept constant
should collapse when  plotted against the scaled distance $ x/\ell_G(t) \simeq x / \sqrt{ t} $, since $z_d=2$ in the voter model. 
In order to put this proposal to the test
we computed the correlation function on square lattices with sizes $ L_i = 2^i L_0 $ for $ L_0 = 80 $ and $ i= 0 - 4$
and times $ t_i = 2^{ i \cdot \alpha_P}  t_0 $, with $ t_0 =256 $, such that $\chi= t_0^{1/\alpha_P}/L_0$.
As far as the exponent $ \alpha_P $ is concerned, we estimated it from the analysis of the largest cluster 
obtaining $ \alpha_P \approx 1.667 $, see Sec.~\ref{subsec:largest-cluster}, and we confirmed its value with the 
study of the time evolution
of the number of percolating domains, see Sec.~\ref{subsec:wrapping_domains}. \par
\begin{figure}[h]
\centering
        \subfloat[]{%
	        \centering
                \includegraphics[scale=0.6]{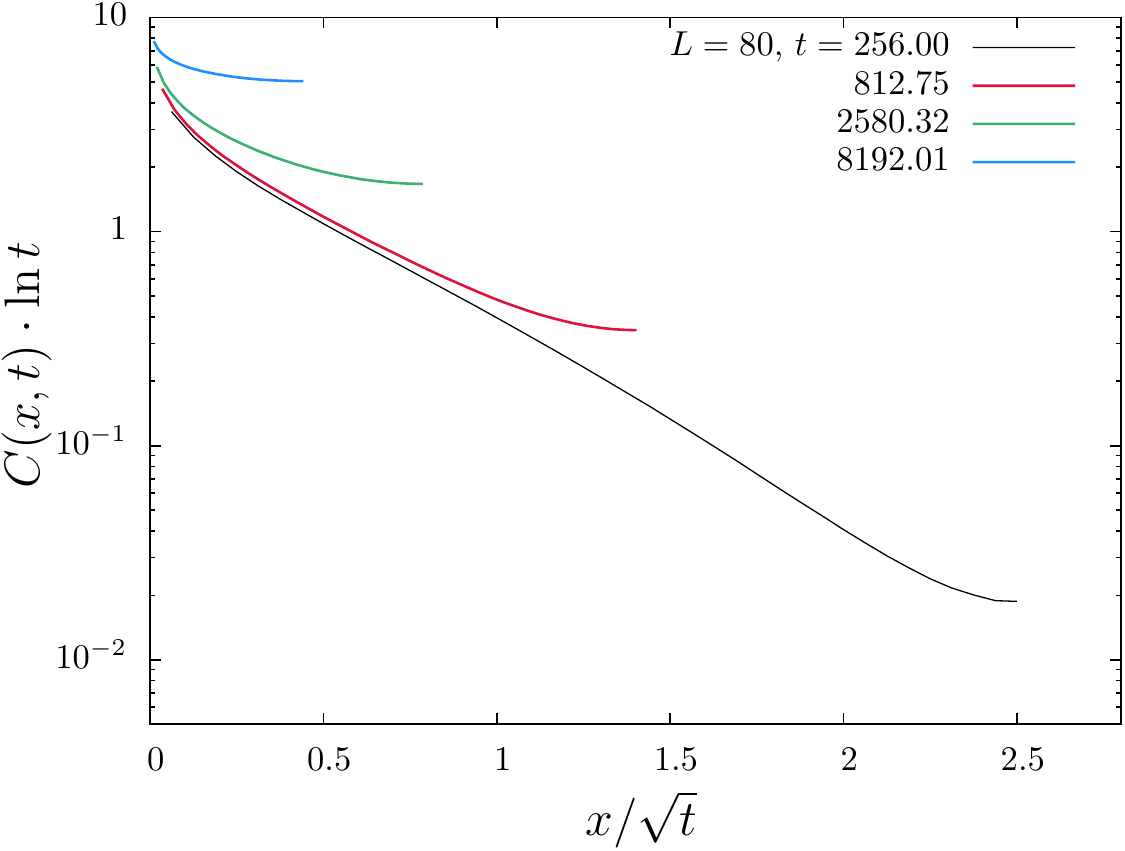}
	        \label{A}
	}%
	\qquad
      	\subfloat[]{%
                \centering
                \includegraphics[scale=0.6]{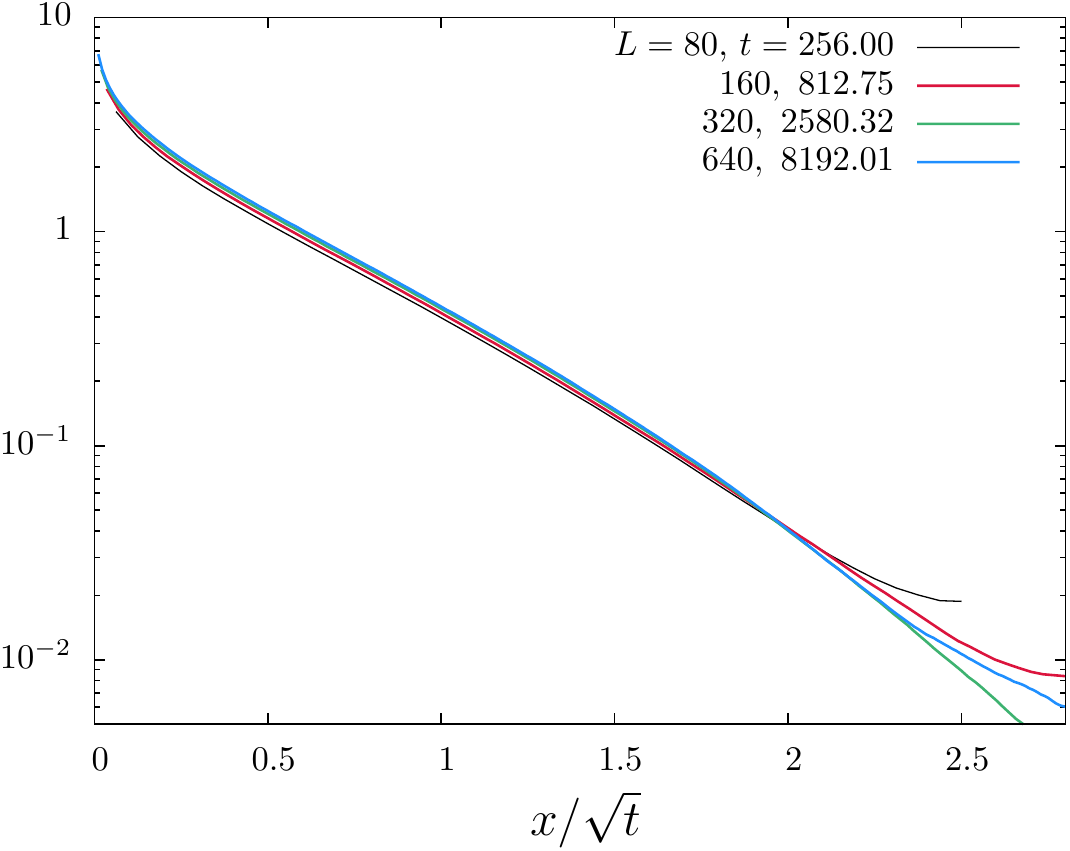}
	        \label{B}
	}%
        \caption{\footnotesize{$ C(x,t) \cdot \ln{t} $ against scaled distance $ x /t^{1/z_d} $ with $ z_d=2 $ the dynamical exponent, 
        in a linear-log plot. In panel (a) data are for $ L=80 $ at different times given in the key. Panel (b) reports data
        for different $ L_i $ and the corresponding  times $ t_i $ such that $L^{\alpha_P}_i/t_i $ is held constant. 
        $ \alpha_P \approx 1.667 $ is the exponent in the relation $ t_P \sim L^{\alpha_P} $, with  $ t_P $  the time
        needed for stable percolating domains to establish (see Sec.~\ref{subsec:largest-cluster} for more details on $t_P$).
}}
\label{fig:CF_Lscaling}
\end{figure}

Figure~\ref{fig:CF_Lscaling} (a) and (b) presents the scaling forms in Eqs.~(\ref{correlf_asympt3})
and~(\ref{CF_new_scaling_form}), respectively. It is clear that the introduction of an extra scaling 
variable with the dependence on the new length scale allows us to achieve a much better data collapse.

\section{Conclusions}
\label{sec:conclusions}

The main goal of this work was to improve the understanding of coarsening in models with microscopic dynamics 
that are not driven by the minimisation of a thermodynamic potential and do not satisfy detailed 
balance. More precisely, we focused on a ${\mathbb Z}_2$ symmetric lattice model with pairwise interactions
driven by interfacial noise, {\it viz.} the $2d$ linear voter model on a square lattice.

We showed that the dynamic evolution of the bidimensional voter model on a square lattice 
proceeds in two distinct dynamic regimes. In the first one, the model approaches critical 
site percolation. The time needed to reach one such typical state diverges with the size of the 
system algebraically, $t_P \simeq L^{1/\alpha_P}$, with the exponent $\alpha_P\simeq 1.667$
that is much larger than the one previously evaluated in the $2d$IM quenched to $T=0$~\cite{Blanchard}. 
Next, the model evolves
following a different mechanism in which consensus is progressively attained. The characteristic
growing length for this process is also algebraic, $\ell(t) \simeq t^{1/z_d}$, though with a different dynamic exponent, $z_d=2$.
In the social dynamics context, the first process can have important consequences.

We based the conclusions above on the careful use of numerical methods. We first tested 
this approach  against the theoretical predictions that were already available 
for infinite size voter model. Most of the computed quantities, such as the fraction of active interfaces, the autocorrelation 
function and the persistence, were found to be in very good agreement with the analytic predictions for infinite size systems.
In particular, the peculiar logarithmic decay of the magnitude of the two-body correlation function and of the fraction of active interfaces
was recovered, even though these results could be improved  by simulating larger systems on very long times. We then focused 
on the spin configurations and from the analysis of their statistical and geometric 
properties we uncovered the approach to critical   site percolation. Once the new 
growing length scale identified, we used it to improve the scaling of the space-time correlation
function for finite size systems.

In a series of papers, the role played by the approach to critical percolation in spin models with 
Ising~\cite{ArBrCuSi07,SiArBrCu07,Blanchard,BlCoCuPi14,SiArBrCu08,SiSaArBrCu09} or Potts~\cite{LoArCuSi10,LoArCu12} variables 
in two or three dimensions~\cite{ArCuPi15} was studied. In all these cases
the dynamics satisfy detailed balance and eventually take a finite size system to thermal equilibrium.
In this paper we explored a different kind of microscopic dynamics that does not satisfy detailed balance 
and approaches an absorbing state asymptotically. We still found a similar approach to critical percolation as in 
the `equilibrium' cases albeit with a much slower growing length.

In~\cite{BlCoCuPi14} we observed that, for the Ising model with microscopic dynamics satisfying detailed balance 
the exponent $\alpha_P$ coincided with the ratio between 
the dynamic exponent for the late stage growth, $z_d$, and the lattice regular or averaged 
coordination number, $n_c$, {\it i.e.} $\alpha_P=z_d/n_c$. In the voter model the dynamic exponent is $z_d=2$ (on top, 
dynamic scaling for infinite size systems suffers from logarithmic corrections)
and, though a coordination number cannot be really identified, we can claim that an effective one is 
somehow larger than one. With this identification, the value of $\alpha_P$ found numerically has the good trend, in the 
sense that the coordination number is smaller than in Ising and this leads to a larger $\alpha_P$.

This works opens several lines for future research. On the one hand, 
it would be interesting to extend this analysis
to different types of lattices, variants of the update rule (with, e.g., local conservation laws~\cite{Caccioli13}, 
memory~\cite{Stark08} or inhomogeneities in the form of zealots~\cite{Mobilia03}) 
and upgrading the voters to have many opinions  (see~\cite{Starnini12} and references therein). 
On the other hand, it should be possible to extract the growing length $\ell_P(t)$ analytically by 
taking into account the finite size effects in the approach explained in Sec.~\ref{sec:analytic}.

\vspace{1cm}

\noindent
{\bf Acknowledgements.}
 L.~F.~C. is a member of Institut Universitaire de France.

\vspace{1cm}

\section*{References}
\bibliographystyle{iopart-num}
\bibliography{voter}

\end{document}